\documentclass[fleqn,usenatbib]{mnras}

\usepackage{newtxtext,newtxmath}

\usepackage[T1]{fontenc}
\usepackage{ae,aecompl}

\usepackage{bm}
\usepackage{graphicx}	
\usepackage{amsmath}	
\usepackage{adjustbox}
\usepackage{comment}
\usepackage{siunitx}
\usepackage{float}
\usepackage[table,xcdraw]{xcolor}
\usepackage{xspace}
\usepackage{etoolbox}
\usepackage{multirow}

\defcitealias{Wegg19}{W19}
\newcommand {\WG} {\citetalias{Wegg19}\xspace}

\newcommand {\kms} {\,{\rm km\,s}^{-1}}

\newcommand {\kpc} {\,{\rm kpc}}

\newcommand {\Gyr} {\,{\rm Gyr}}

\newcommand{\te}[1]{\text{#1}}
\newcommand{\gaia}{\textit{Gaia} }
\newcommand{\gaiap}{\textit{Gaia}}

\definecolor{mygreen}{rgb}{0.0, 0.42, 0.24}

\title[Chemo-kinematics of the {\it Gaia} RR Lyrae]{Chemo-kinematics
  of the {\it Gaia} RR Lyrae: the halo and the disc}

\author[G. Iorio and V. Belokurov]{
Giuliano Iorio$^{1,2,3}$\thanks{giuliano.iorio.astro@gmail.com} and
Vasily Belokurov$^{3}$\thanks{vasily@ast.cam.ac.uk}
\\
$^{1}$Physics and Astronomy Department Galileo Galilei, University of Padova, Vicolo dell'Osservatorio 3, I--35122, Padova, Italy\\
$^{2}$INFN - Padova, Via Marzolo 8, I--35131 Padova, Italy\\
$^{3}$Institute of Astronomy, University of Cambridge, Madingley Road, Cambridge CB3 0HA, UK\\
}

\date{Accepted XXX. Received YYY; in original form ZZZ}

\pubyear{2020}

\begin{document}
\label{firstpage}
\pagerange{\pageref{firstpage}--\pageref{lastpage}}
\maketitle

\begin{abstract}
We present the results of a multi-component kinematic model of a large
sample of RR Lyrae detected by \gaiap. By imposing a four-fold
symmetry and employing \gaia proper motions, we are able to infer the
behaviour of the velocity ellipsoid between $\approx3$ and $\approx30$
kpc from the centre of the Galaxy.  We detect the presence of two
distinct components: a dominant non-rotating halo-like population and
a much smaller rotating disc-like population.  We demonstrate that the
halo RR Lyrae can be described as a superposition of an isotropic and
radially-biased parts. The radially-biased portion of the halo is
characterised by a high orbital anisotropy $\beta\approx0.9$ and
contributes between 50\% and 80\% of the halo RR Lyrae at
$5<R$(kpc)$<25$. In line with previous studies, we interpret this
high-$\beta$ component as the debris cloud of the ancient massive
merger also known as the {\it Gaia} Sausage (GS) whose orbital extrema
we constrain. The lightcurve properties of the RR Lyrae support the
kinematic decomposition: the GS stars are more metal-rich and boast
higher fractions of Oosterhoff Type 1 and high amplitude short period
(HASP) variables compared to the isotropic halo component. The
metallicity/HASP maps reveal that the inner 10 kpc of the halo is
likely inhabited by the RR Lyrae born in-situ. The mean azimuthal
speed and the velocity dispersion of the disc RR Lyrae out to
$R\approx30$ kpc are consistent with the behaviour of a young and
metal-rich thin disc stellar population.

\end{abstract}




\section{Introduction}

The simple and convenient picture in which the Galaxy is made up of
clear-cut structural blocks, largely independent yet arranged to work
in concert, is falling apart before our eyes. The harbinger of this
paradigm shift is the mushrooming of dualities -- today every piece of
the Milky Way has acquired a sidekick: there are two discs, `thin'
and `thick' \citep[or more precisely, $\alpha$-poor and
  $\alpha$-rich,
  see][]{Gilmore1983,Fuhrmann1998,Bensby2003,Haywood2008,Bovy2012,Hayden2015},
the accreted halo must be distinguished from the one built in-situ
\citep[e.g.][]{Searle1978,Helmi1999,Brook2003,Venn2004,Bell2008,Nissen2010,Bonaca2017,Gallart2019,Belokurov2020},
and the bulge is really a bar, or perhaps several
\citep[][]{Blitz1991,Binney1997,Zoccali2003,McWilliam2010,Robin2012,Ness2013,Wegg2013,Bensby2013}.

Thanks to the ESA's {\it Gaia} space observatory
\citep[][]{Prusti2016} we are reminded that, in fact, the Galaxy is an
evolving and interconnected system where components may interact and
can profoundly affect each other. For instance, it is now clear that
the last significant merger that formed the bulk of the stellar halo
\citep[][]{Deason2013,Belokurov2018,Haywood2018,Helmi2018,Mackereth2019,Fattahi2019}
may be connected to a series of metamorphoses occurring in the young
Milky Way. This early accretion event revealed by the unprecedented
astrometry from {\it Gaia} not only dictates the structure of the
inner stellar halo
\citep[][]{Deason2018,HaloAction,SausageGCs,Koppelman2018,Lancaster2019,Iorio19,Simion2019,Bird2019}
but appears to be contemporaneous with the demise of the thick disc,
emergence of the in-situ halo and the formation of the bar
\citep[][]{DiMatteo2019,Fantin2019,Belokurov2020,Grand2020,Bonaca2020,Fragkoudi2020,Sit2020}. These
tumultuous transmutations are not exclusive to the Galaxy's youth --
signs have been uncovered of the ongoing interactions quaking the
Galactic plane \citep[][]{Minchev2009,Widrow2012,Xu2015}, including
pieces of evidence procured recently using the {\it Gaia} Data Release
2 \citep[see][]{Antoja2018,Laporte2019,Bland-Hawthorn2019}. Even today
it is easy to start in the disc and end up in the halo
\citep[][]{Michel-Dansac2011,Price-Whelan2015,Gomez2016,JB2017,Laporte2018,deBoer2018}.

In this time of confusion, reliable distance and age/metallicity
indicators are essential to building a coherent picture of the Milky
Way. For decades, pulsating horizontal branch stars known as RR Lyrae
(RRL, hereafter) have been trusted upon to help us chart the Galaxy
\citep[e.g.][]{Kinman1966,Oort1975,Saha1985,Hartwick1987,Catelan2009,Pietrukowicz2015}.
Using painstakingly-assembled spectroscopic samples it has been
established that RRL metallicities span a wide range but the stars
appear predominantly metal-poor, while the analysis of the Galactic
Globular clusters revealed prevalence for old ages
\citep[][]{Preston1959,Butler1975,Sandage1982,Suntzeff1991,Lee1994,Clementini1995,Clement2001}. Note
that in the field, RRL are sufficiently rare, therefore no large
spectroscopic datasets are currently available. However, an
approximate metallicity estimate can be gauged from the properties of
the lightcurve alone \citep[][]{Sandage1982,
  Carney1992,Nemec1994,Jurcsik96,Nemec13}.

\begin{figure*}
\centering
\centerline{\includegraphics[width=1.0\textwidth]{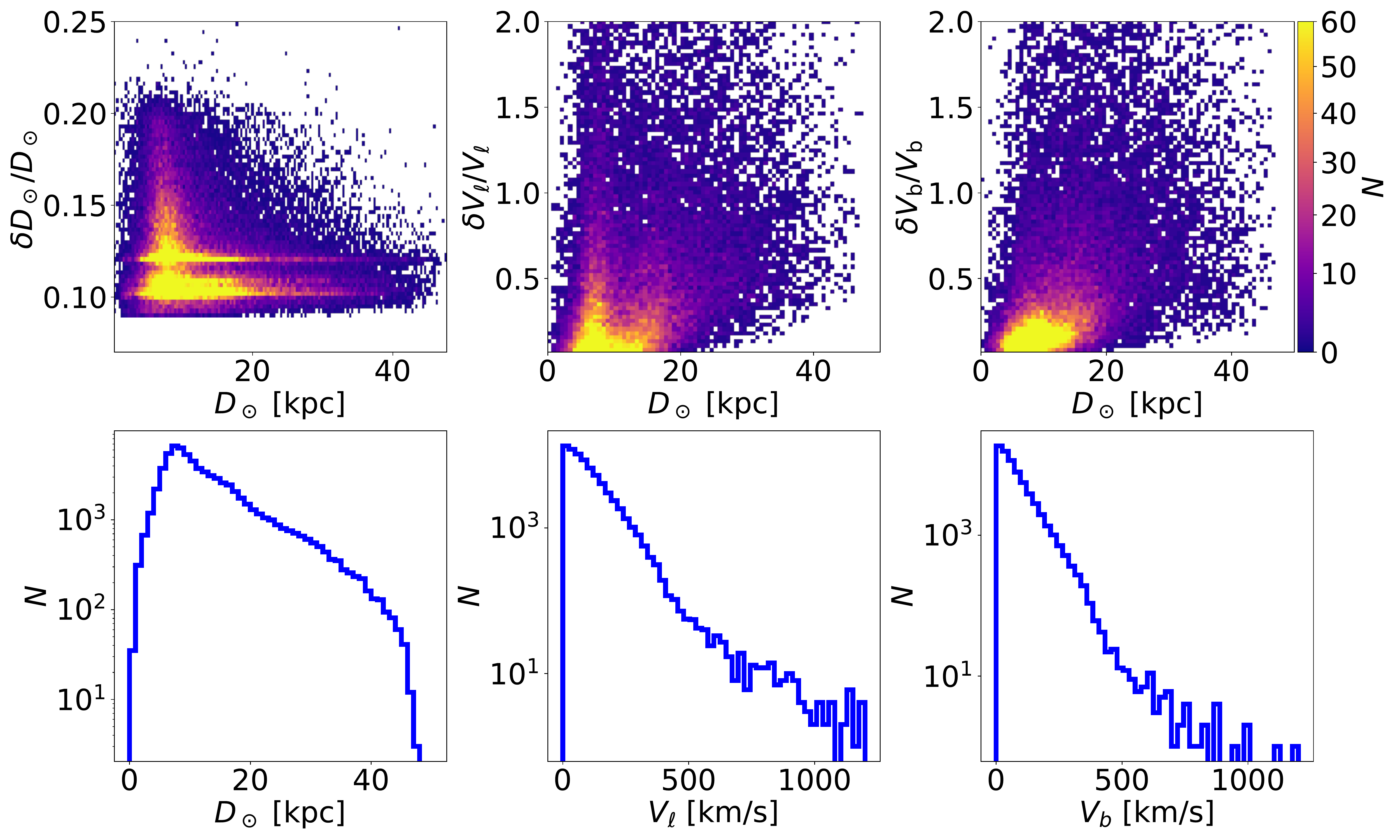}}
\caption[]{Distances and transverse velocities for stars in the Gclean
  catalogue (see Section~\ref{sec:cleaning}). Top panels show density
  distributions in the plane of relative error (absolute value) and
  heliocentric distance. Bottom panels give distributions of the
  heliocentric distance and the components of the apparent
  (sky-projected) tangential velocity. Note that this plot also shows
  stars with $D_\odot>40 \ \kpc$ that are eliminated in the final
  Gclean catalogue.}
\label{fig:hist_sample}
\end{figure*}

In the last two decades, wide-area multi-epoch surveys have brought in
a rich harvest of variable stars in general and RRL in particular
\citep[e.g.][]{Sesar2007,Soszynski2009,Drake2013,Soszynski2014,Torrealba2015,Sesar2017}. Typically
old and metal-poor, RRL have long served as a tried and true
tracer of the Galactic halo and its sub-structures
\citep[e.g.][]{Vivas2001,Morrison2009,Watkins2009,Sesar2013,Simion2014,Mateu18,Hernitschek2018}. \gaiap,
the first truly all-sky variability census in the optical, has further
improved our understanding of the Milky Way RRL,  not only by
filling in the gaps left behind by the previous generations of
surveys, but also by providing high-quality proper motions for the
bulk of the RRL  it sees. The \gaia data has thus enabled a new,
precise characterisation of the Galactic halo density field
\citep[e.g.][]{Iorio18,Wegg19,Iorio19} and helped to discover halo
sub-structures previously not seen
\citep[][]{Belokurov2017,Koposov2019,Pisces2019,Torrealba2019}.

While it is true that RRL  are being used primarily to trace the
fossil record of the Milky Way assembly, it was always known that in
the field, a relatively small number of metal-rich examples exist
\citep[][]{Kukarkin1949,Preston1959,Smith84,Layden94,Walker91,Dekany18,Chadid2017,Fabrizio2019,Zinn20}. 
Based on their
kinematics, these metal-rich RRL  were assigned to the Galactic
disc(s) \citep[][]{Layden1995a}. Given the enormous number of
available red giant progenitors, metal-rich RRL  in the disc were
estimated to form between 200 and 800 times less often compared to
their old and metal-poor halo counterparts
\citep[][]{Taam1976,Layden1995b}. While the formation channel has not
yet been identified, these early studies as well as the subsequent
follow-up conjectured that the progenitors of metal-rich RRL 
ought to be old, i.e. $>10$ Gyr \citep[e.g.][]{MateuDisc}. 
{
The presence of likely old metal-rich RRL has been confirmed also in metal-rich Globular Clusters (e.g.\ NGC 6338 and NGC 6441, see \citealt{Pritzl2000}), however  they have periods that are significantly larger  with respect to field metal-rich RRL.}
The main
obstacle to the production of a metal-rich RRL  is its temperature
on the HB: with higher envelope opacities, these stars tend to sit too
far to the red from the instability strip
\citep[e.g.][]{Dorman1992}. Therefore, before arriving onto the HB,
metal-rich RRL  progenitors are required to undergo copious levels
of mass-loss, $\approx0.5$ M$_{\odot}$ or more, which may well be
beyond what is physically possible.

Most recently, the conundrum of metal-rich RRL  has been given a
new lease of life. \citet{Marsakov18} demonstrated that while plenty
of the local metal-rich RRL  likely belong to the thick disc (and
thus can be as old as $\approx$10 Gyr), a substantial fraction
displays the kinematics of the younger portion of the thin disc. An
age of only few Gyrs would be very difficult to reconcile with the
conventional scenarios of the RRL  formation. Note that if extreme
mass loss can be invoked, i.e. in excess of $1$ M$_{\odot}$, then even
young ($>1$ Gyr) progenitors can produce metal-rich RRL 
\citep[see][]{Bono1997,Bono97b}. In a follow-up study, \citet{Marsakov2019}
estimated the masses of the metal-rich thin disc RRL  and found
them to be of order of $0.5-0.6$ M$_{\odot}$, thus confirming the need
for mass loss beyond the typically accepted values. Finally,
\citet{Zinn20} and \citet{Prudil2020} combined RRL  with available
spectroscopy with the {\it Gaia} DR2 astrometry to confirm the
existence of metal-rich RRL  stars with the orbital properties
typical of the Galactic thin disc. With these most recent observations
in hand, it remains to be seen if metal-rich RRL  can actually be
easily accommodated within the current stellar evolution theory. Comparing the
structural properties of the metal-rich and metal-poor RRL, 
\citet{Chadid2017} conclude that it can not.

What is hard to achieve via single stellar evolution channels can
(sometimes) be effortlessly done with binary stars. Indeed, an object
has been discovered that nimbly mimics the classic RR Lyrae behaviour,
i.e.  lives on the instability strip and pulsates with the same kind
of lightcurve, yet it is not an RR Lyrae, at least not in the
conventional meaning of the term \citep[][]{BEP}. This star,
designated Binary Evolution Pulsator (BEP), is a low-mass
($0.26 \ M_{\odot}$) remnant of mass transfer in a binary system with a
period of $\approx15$ days. As the follow-up theoretical work
demonstrates, binary evolution can lead to a broad range of BEP
masses, and in some cases even involve a stripped star with a
helium-burning core \citep[][]{BEPoccurrence}. These impostors would
be indistinguishable from the classic RR Lyrae but have an age of only
4-5 Gyr. Only one such object has been found so far, but searches for
RR Lyrae in binary systems are ongoing \citep[e.g.][]{Prudil2019,
  Kervella2019}.

This work aims to exploit the unprecedented all-sky coverage of \gaia
to study the chemo-kinematics of the halo and the disc of the Milky
Way as traced by RRL stars. The paper is organised as
follows. Section~\ref{sec:sample} presents the construction of a clean
sample of \gaia RRL stars and gives the details of the methods we use
to estimate physical quantities like distance, metallicity and
transverse velocity. Section~\ref{sec:method} describes the machinery
employed to perform the kinematic decomposition of the Galactic
components. Then, we discuss the properties of the individual
components: the halo in Section~\ref{sec:halo} and the disc in Section~\ref{sec:disc}. In
Section~\ref{sec:summary} we discuss possible biases affecting the
results and finally, we summarise the main conclusions.
  
\section{The sample} \label{sec:sample}

We use the whole catalogue of stars classified as RRL in \gaia DR2
\citep{GaiaDR2} combining the SOS (Specific Object Study,
\citealt{Clementini}) RRL catalogue with the stars classified as RRL in the
general variability table \texttt{vari\_classifier\_result} \citep{GaiaVariable}
following the procedure described in \cite{Iorio19}.  The initial
combined catalogue contains 228,853 stars ($\approx77 \%$ RRab,
$\approx21 \%$ RRc and $\approx2 \%$ RRd).

\subsection{Distance and velocities estimate} \label{sec:dist_estimate}

One of the key ingredients of this analysis is the distance from the
Sun, $D_\odot$, of each star. Once the heliocentric distance is known,
we estimate the Galactocentric coordinates and, using the observed
proper motion, calculate the velocities $V_\ell$ (along the Galactic
longitude $\ell$) and $V_b$ (along the Galactic latitude).

\noindent
{\bf Galactic parameters.}  We set a left-handed Galactocentric frame
of reference similar to the one defined in \cite{Iorio18}: here
$x$,$y$,$z$ indicate the Cartesian coordinates; $R$ is the cylindrical
radius, $r$ is the spherical radius and $\phi$, $\theta$ represent the
azimuthal and zenithal angle. In this coordinate system the Sun is
located at $x_{\odot}=R_{\odot}=8.13\pm0.3$ kpc \citep{Rsun} and
$z_{\odot}=0$ kpc (see \citealt{Iorio18}). In order to correct the
observed stellar velocity for Sun's motion, we adopt
$V_\mathrm{lsr}=238\pm9 \ \kms$ \citep{Vlsr} for the local standard of
rest (lsr) and $(U_\odot,V_\odot,W_\odot)=(-11.10\pm1.23,
12.24\pm2.05,7.25\pm0.63) \ \kms$ \citep{UVWsun} for the Sun's proper
motion with respect to the lsr (assuming the Galactocentric frame of
reference defined above).  The final correcting vector is
\begin{equation}
V_\mathrm{\odot,corr}=(-11.10\pm1.23, 250.24\pm9.23, 7.25\pm0.63).
\label{eq:Vcorr}
\end{equation}

In order to take into account all of the uncertainties in the estimate
of the physical parameters of interest, we use a Monte-Carlo sampling
method ($10^5$ realisations) following the steps: i)
correction of \gaia $G$ magnitudes for the dust reddening, $ii)$
estimate of the metallicity, $iii)$ estimate of the absolute magnitude
$M_\mathrm{G}$, $iv)$ estimate of the distance and the Galactocentric
coordinates, $v)$, estimate of the velocities. Where not specified we
sample the value of a given parameter $X=\bar{X} \pm \delta X$ drawing
variates from a normal distribution centred on $\bar{X}$ and with a
standard deviation $\delta X$.

\begin{figure}
\centering
\centerline{\includegraphics[trim=13cm 0cm 13cm 0cm,clip=true,width=1.0\columnwidth]{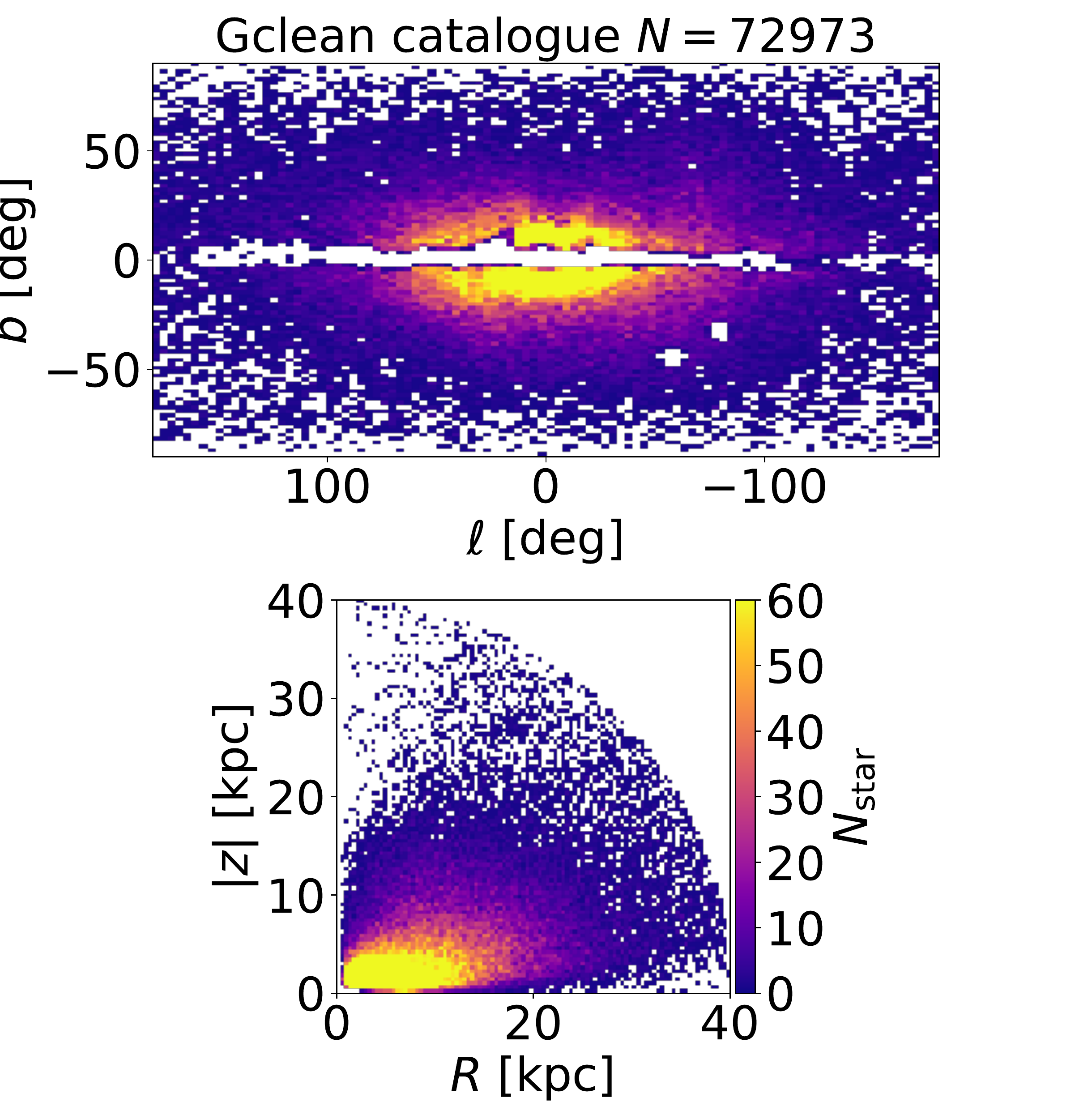}}
\caption[]{Galactic (top panel) and Galactocentric cylindrical (bottom
  panel) star count maps for objects in the Gclean catalogue (see
  Section~\ref{sec:cleaning})}
\label{fig:map_total}
\end{figure}
\bigskip
\noindent
{\bf Magnitude correction for dust reddening.}  We correct the
observed $G_\mathrm{obs}$ magnitude as
\begin{equation}
G=G_\mathrm{obs}-k_\mathrm{G} E(B-V),
\label{eq:Gcorr}
\end{equation}
where $E(B-V)$ and its error, $\delta_\mathrm{E(B-V)}=0.16\times
E(B-V)$, comes from \cite{dustext}.  The factor $k_\mathrm{G}$ is
obtained by applying Equation 1 of \cite{Babusiaux18} iteratively if
the star has an estimate of the \gaia color $BP-RP$, otherwise we
assume $k_\mathrm{G}=2.27 \pm0.30$ \citep{Iorio19}.  
{For the stars in 
the SOS catalogue, the adopted $G_\mathrm{obs}$ is the SOS table entry
\texttt{int\_average\_g} and the color $BP-RP$ is the difference between the columns 
\texttt{int\_average\_bp} and \texttt{int\_average\_rp}. For the other stars, we use the values reported in the general  \gaia source catalogue (\texttt{phot\_g\_mean\_mag}, \texttt{phot\_bp\_mean\_mag}, \texttt{phot\_rp\_mean\_mag}).
We notice a small  offset ($\approx0.03$  for $G_\mathrm{obs}$ and $\approx0.02$ for $BP-RP$) between the SOS and general \gaia values, hence we correct the latter.  We use the  values from the
SOS catalogue as standard for two reasons: they are estimated directly from the lightcurves (robust against outliers, see \citealt{Clementini})  and the magnitude-metallicity relation we use (see below) has been calibrated on these $G$ values
(see \citealt{Muraveva18}).
After the offset correction, the differences between the SOS and \gaia observed magnitudes can be treated as another source of random errors on the estimate of $G$. For most of the stars in the sample (> 98 \%) the magnitude of this error is $\lesssim0.1\%$, representing a negligible amount in the error budget of the final distance estimate (see below). We decided to not  consider the errors on $G_\mathrm{obs}$, thus the error on $G$ comes only from the uncertainties on  $k_\mathrm{G}$ or $E(B-V)$.}

\noindent
{\bf Metallicity estimate.}  It is well known that the metallicities
of RRL correlate with their lightcurve properties
(e.g.\ \citealt{Jurcsik96,Smolec05,Nemec13,Dekany18fit}). Two of the
most used properties are the period (fundamental period, $P$ for RRab
stars, first overtone period, $P_\mathrm{1o}$, for RRc stars) and the
phase difference between the third and the first harmonics $\Phi_{31}$
of the lightcurve decomposition. Although the SOS catalogue already
reports an estimate of the metallicity based on the \citealt{Nemec13}
relations (see \citealt{Clementini}), we decide to use instead a
linear relation calibrated directly on the \gaia $P$ (or
$P_\mathrm{1o}$) and $\Phi_{31}$ parameters (see
e.g.\ \citealt{Jurcsik96}).  For the RRab stars we cross-match the SOS
catalogue with the spectroscopic sample of \cite{Layden94} finding 84
stars in common and deriving the following relation:
\begin{equation}
\begin{aligned}
\mathrm{[Fe/H]}_\mathrm{RRab}= & (-1.68\pm0.05) + (-5.08\pm0.5)\times(P-0.6) \\
                & + (0.68\pm0.11)\times (\Phi_{31}-2.0), 
\end{aligned}
\label{eq:Met_RRab}
\end{equation}
with an intrinsic scatter $\delta_{\mathrm{[Fe/H]}}=0.31\pm0.03$.
Concerning the RRc, following \cite{Nemec13}, we use the RRc stars in
known Globular Clusters as classified by \cite{HelmiGaia}, then we
assign to each of them the metallicity reported for the Globular
Clusters in
\cite{Globularcluster}\footnote{\url{http://vizier.u-strasbg.fr/viz-bin/VizieR?-source=VII/202}}. Using
this method we obtain the following metallicity relation:
\begin{equation}
\begin{aligned}
\mathrm{[Fe/H]_\mathrm{RRc}}= & (-1.26\pm0.03) + (-9.39\pm0.66)\times(P_\mathrm{1o}-0.3) \\                & + (0.29\pm0.05)\times(\Phi_{31}-3.5), 
\end{aligned}
\label{eq:Met_c}
\end{equation}
with an intrinsic scatter $\delta_\mathrm{[Fe/H]}=0.16\pm0.03$.  We
sample the metallicity distribution for each star drawing from both
the $P$ (or $P_\mathrm{1o}$) and $\Phi_{31}$ distributions considering
their errors and from the posterior of the model parameters (taking
into account their correlation).  In case the star has not a period
estimate and/or $\Phi_{31}$, these values are drawn from their overall
2D distribution considering the whole \gaia SOS catalogue. After this
step we end up with $10^5$ [Fe/H] realisations for each star.  Further
information on the metallicity estimate can be found in the Appendix
\ref{appendix:met}.

\begin{figure*}
\centering
\centerline{\includegraphics[width=1.1\textwidth]{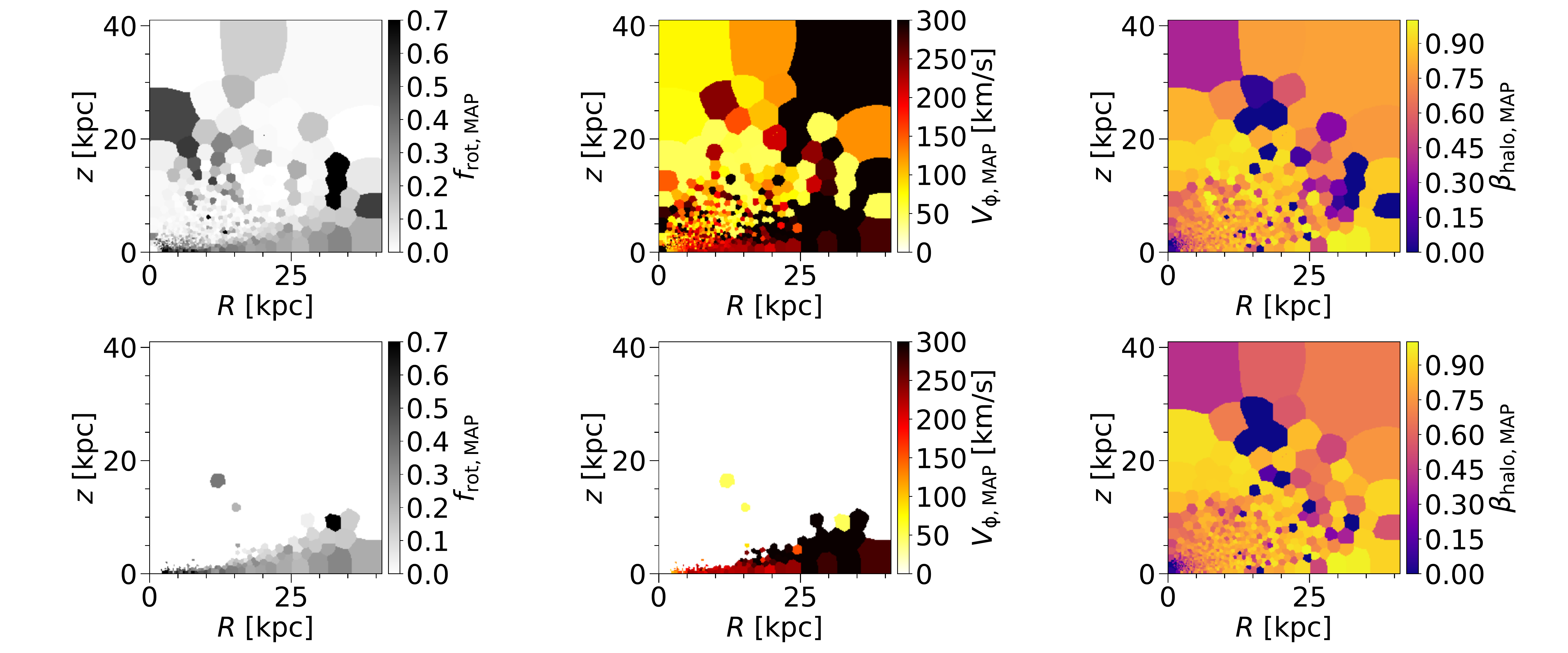}}
\caption[]{{\it Top:} results of the double-component fit for the RRLs
  in the Gclean sample (see Table~\ref{tab:kseparation}).
  Maximum-a-posteriori (MAP) values are shown in cylindrical
  coordinates.  {\it Bottom}: mixed 1 and 2 component results (results
  from the double-component fit if $\Delta \mathrm{BIC}>10$, otherwise
  from the single-component fit, see text for details). Left column
  shows the fraction of the rotating component, middle panels give the
  azimuthal velocity of the rotating component, while right column
  presents the anisotropy of the halo-like component.}
\label{fig:disc_halo}
\end{figure*}

\noindent
{\bf Absolute magnitude.} The absolute magnitudes are estimated using
the $M_\te{G}-\mathrm{[Fe/H]}$ relation described in
\cite{Muraveva18}. We sample the absolute magnitude distribution for
each star using the $\mathrm{[Fe/H]}$ realisations (see above) and
drawing the $M_\te{G}-\mathrm{[Fe/H]}$ relation parameters (taking
into account the intrinsic scatter) using the errors reported by
\cite{Muraveva18}.

\noindent
{\bf Distance estimate.} We produce $10^5$ realisations of the
heliocentric distance using the familiar equation
\begin{equation}
\log\left(\frac{D_\odot}{\te{kpc}}\right)=\frac{G-M_\mathrm{G}}{5}-2.
\label{eq:amag}
\end{equation}
\noindent Then, the heliocentric distance and the observed Galactic
coordinates ($\ell$, $b$, taken without their associated
uncertainties) are used to obtain realisations of the Galactocentric
Cartesian, cylindrical and spherical coordinates ($x$,$y$,$z$,$R$,$r$,$\phi$,$\theta$) taking into account the errors on the Galactic
parameters.  Finally, we use the mean and the standard deviation of
the final realisations to obtain the fiducial value and errors on the
Galactic coordinates for each star.

\noindent
{\bf Velocity estimate.} We estimate the physical velocities from the
observed proper motions as
\begin{equation}
\begin{aligned}
V_\ell &= K\mu_\ell D_\odot + V_{\ell,\odot} \\ 
V_\mathrm{b} &=K\mu_b D_\odot  + V_\mathrm{b,\odot}
\end{aligned}
\label{eq:Vlb}
\end{equation}
where $K \approx4.74$ is the conversion factor from $\te{mas}
\ \te{kpc} \ \te{yr}^{-1}$ to $\te{km}
\ \te{s}^{-1}$. $V_{\ell,\odot}$ and $V_\mathrm{b,\odot}$ represent
the projection of the Sun velocity (Equation~\ref{eq:Vcorr}) in the
tangential plane at the position of the star. These two values are
estimated by applying the projection matrix defined in Equation A2 in
\cite{Iorio20} to the correcting vector in Equation~\ref{eq:Vcorr}.
We draw $10^5$ realisations for each star taking into account the
$D_\odot$ samples, the errors and the covariances of the proper
motions and the errors on $V_\mathrm{\odot,corr}$.  Then, we estimate
the mean value, the standard deviation and the covariance between
$V_\ell$ and $V_\mathrm{b}$. We use these values to perform our
kinematic analysis (see Section~ \ref{sec:method}).

\subsection{Cleaning} \label{sec:cleaning}

In order to study the global properties of the (large-scale) Galactic
components, we clean the RRL sample by removing the stars belonging to
the most obvious compact structures (Globular Clusters and dwarf
galaxies including the Magellanic Clouds) as well as various artefacts
and contaminants.  This procedure is similar to the cleaning process
described in \cite{Iorio19}, especially with regards to the cull of
known Galactic sub-structures. Concerning the artefacts and
contaminants, we employ a slightly different scheme in order to
both maintain as many stars at low latitudes as possible and have more
robust quality cuts. In particular, we focus on removing stars that
could have biased astrometric solutions or unreliable photometry.
\bigskip

\noindent
{\bf Artefacts and contaminants.} \cite{GaiaVariable},
\cite{Clementini} and \cite{Rimoldini19} found that in certain regions
(the bulge and the area close to the Galactic plane) the presence of
artefacts and spurious contaminants in the \gaiap's RRL catalogues can
be quite significant. The contaminants in these crowded fields are
predominantly eclipsing binaries and blended sources, with a minute
number of spurious defections due to misclassified variable stars
\citep{GaiaVariable}. To remove the majority of the likely
contaminants we apply the following selection cuts:

\begin{itemize}
    \item $RUWE$<1.2
    \item $1.0 + 0.015\times(BP-RP)^2<BRE<1.3+0.06\times(BP-RP)^2$
    \item $E(B-V)$<0.8
\end{itemize}{}

The \texttt{renormalised\_unit\_weight\_error} ($RUWE$) is expected to
be around one for sources whose astrometric measurements are
well-represented by the single-star five-parameter model as described
in \citet{Lindegreen18}.  Therefore the above $RUWE$ cut eliminates
unresolved stellar binaries \citep[see e.g.~][]{Belokurov20} as well
as blends and galaxies \citep[see e.g.~][]{Koposov17}.  The
\texttt{phot\_bp\_rp\_excess\_factor}, $BRE$, represents the ratio
between the combined flux in the \gaia $BP$ and $RP$ bands and the
flux in the $G$ band, and thus by design is large for blended sources
\citep[see][]{EvansGaia}.  Following \cite{Lindegreen18}, we remove
stars with $BRE$ larger or lower than limits that are functions of the
observed colors (Equation~C2 in \citealt{Lindegreen18}).  Finally, we
remove stars in regions with high reddening, $E(B-V)$ \citep[according
  to][]{dustext}, for which the dust extinction correction is likely
unreliable.  After these cuts, our RRL sample contains 115,774 RRL
stars.

\noindent
{\bf Globular clusters and dwarf satellites.} We consider all globular
clusters (GCs) from the \cite{Globularcluster}
catalogue\footnote{\url{http://physwww.mcmaster.ca/~harris/Databases.html}}
and all dwarf galaxies (dWs) from the catalogue published as part of
the \texttt{Python} module
\texttt{galstream}\footnote{\url{https://github.com/cmateu/galstreams}}
(\citealt{Mateu18}). We select all stars within twice the truncation
radius of a GC if this information is present, otherwise we use 10
times the half-light radius. For the dWs we take 15 times the
half-light radius. Amongst the selected objects, we remove only the
stars in the heliocentric distance range $D_\mathrm{GC/dWs}\pm0.25
\times D_\mathrm{GC/dWs}$. The chosen interval should be large enough
to safely take into account the spread due to the uncertainty in the
RRL distance estimate (see Section~\ref{sec:dist_estimate} and
Figure~\ref{fig:hist_sample}).  This procedure removes 1,350 stars.

\noindent
{\bf Sagittarius dwarf.} In order to exclude the core of the
Sagittarius dwarf we select all stars with
$|\tilde{B}-\tilde{B}_\te{Sgr}|<9^\circ$ and $|\tilde{\Lambda} -
\tilde{\Lambda}_\te{Sgr}|<50^\circ$, where $\tilde{B}$ and
$\tilde{\Lambda}$ are the latitude and longitude in the coordinate
system aligned with the Sagittarius stream as defined in
\cite{BelokurovSgr}\footnote{Actually, we use a slightly different
  pole for the Sagittarius stream with $\alpha=303.63^\circ$ (Right
  Ascension) and $\delta=59.58^\circ$ (declination)} and
$\tilde{B}_\te{Sgr}=4.24^\circ$ and
$\tilde{\Lambda}_\te{Sgr}=-1.55^\circ$ represent the position of the
Sagittarius dwarf. Then, among the selected objects, we get rid of all
stars with a proper motion relative to Sagittarius lower than $2
\ \te{mas} \ \te{yr}^{-1}$, considering the dwarf's proper motion from
\cite{HelmiGaia}.  The stars in the tails have been removed
considering all the objects within
$|\tilde{B}-\tilde{B}_\te{Sag}|<11^\circ$ and with proper motions (in
the system aligned with the Sgr stream) within 1.5 mas yr$^{-1}$ from
the proper motions tracks of the Sgr stream (D. Erkal private
communication, the tracks are consistent with the ones showed in
\citealt{Ramos20}). The cuts of the core and tails of the Sgr dwarf
remove 7,233 stars.

\noindent
{\bf Magellanic Clouds.} We apply the same selection cuts as those
used in \cite{Iorio19} thus removing 14,987 stars (11,934 for the LMC
and 3,053 for the SMC).

\noindent
{\bf Cross-match with other catalogues.} In order to identify possible
classification mistakes and other contaminants, we cross-match the
catalogue scrubbed of substructures and artefacts (as described above)
with the $SIMBAD$ astronomical database \citep{simbad}, the $CSS$
periodic variable
table\footnote{\url{http://vizier.u-strasbg.fr/viz-bin/VizieR-3?-source=J/ApJS/213/9/table3&}}
\citep{css} and the
$ASAS$-$SN$\footnote{\url{https://asas-sn.osu.edu/variables}}
catalogue of variable stars \citep{assasn2,assasn3,assasn1}. We remove
all stars that have not been classified as: \texttt{RRLyr},
\texttt{CandidateRRLyr}, \texttt{HB*}, \texttt{Star},
\texttt{Candidate\_HB*}, \texttt{UNKNOWN}, \texttt{V*}, \texttt{V*?}
in $Simbad$ (1,015 stars); \texttt{RRab}, \texttt{RRc} or \texttt{RRd}
in $CSS$ (655 stars) or $ASAS$-$SN$ (11,963 stars).  Analysing these
data we found a low level of contamination (stars not classified as
RRL in the cross-matched catalogue $\lesssim 3 \%$) considering
$Simbad$ and $CSS$, while the level of contamination considering
$ASAS$-$SN$ is ten times larger ($\approx27\%$). However, as most of
the contaminants are classified as \texttt{UNKNOWN} ($\approx20\%$) in
$ASAS$-$SN$, these objects could suffer from poor lightcurve
sampling. Another significant contaminant class is eclipsing binaries,
mostly W Ursae Majoris variables (WUMa, $\approx 5\%$) for which the
lightcurve could be misclassified as an RRc. Indeed, among the stars
classified as WUma in $ASAS$-$SN$ about $80\%$ are classified as RRc
in the \gaia SOS catalogue.  Not considering the dominant sources of
contamination discussed above, the number of unwanted interlopers
estimated from $ASAS$-$SN$ is similar to that obtained with $Simbad$
and $CSS$. Comparing the RRL classification for the stars in common
between the \gaia SOS catalogue and the \gaia general variability
catalogue we decided to remove all stars that have been classified as
RRd (2941 stars) in at least one of the two catalogues. In total these
cuts remove 15,633 stars.


\noindent
{\bf Distance cut.} Given the significant increase in velocity
uncertainties at large distance, we decide to limit the extent of our
sample to within 40 kpc from the Galactic centre. This cut removes
4,057 stars.

\begin{figure}
\centering
\centerline{\includegraphics[width=1.0\columnwidth]{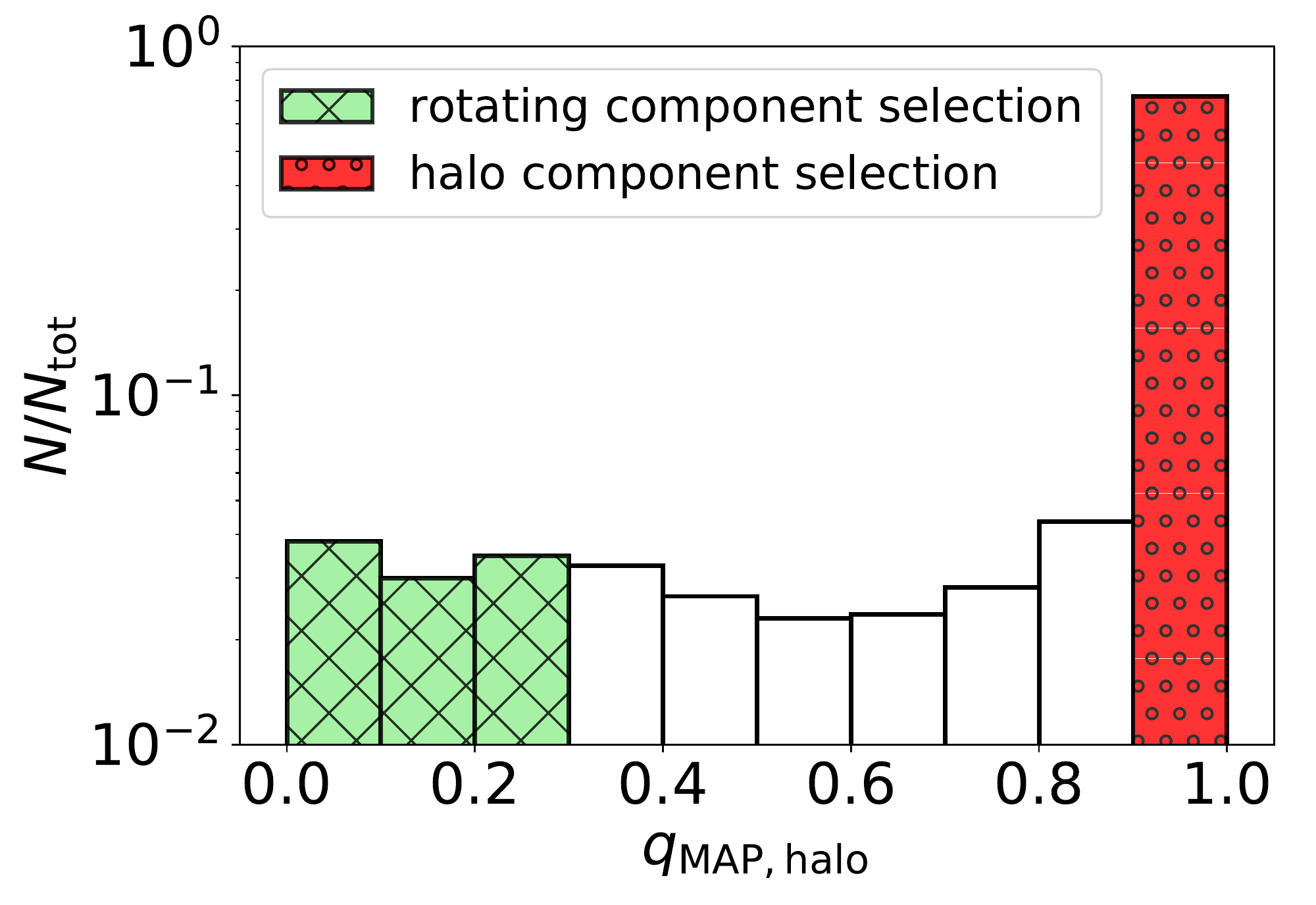}}
\caption[]{Distribution of the RRL maximum-a-posteriori probability
  (MAP, see Section~ \ref{sec:kfit}) of belonging to the non-rotating
  (halo) kinematic component from the double component fit described
  in Section~ \ref{sec:kseparation}. The red o-hatched and the green
  x-hatched regions indicate the $q_{\mathrm{MAP,halo}}$ cuts used to
  select the halo and the rotating (disc-like) subsample
  respectively.}
\label{fig:qhist_halo}
\end{figure}
\begin{figure*}
\centering
\centerline{\includegraphics[width=1.\textwidth]{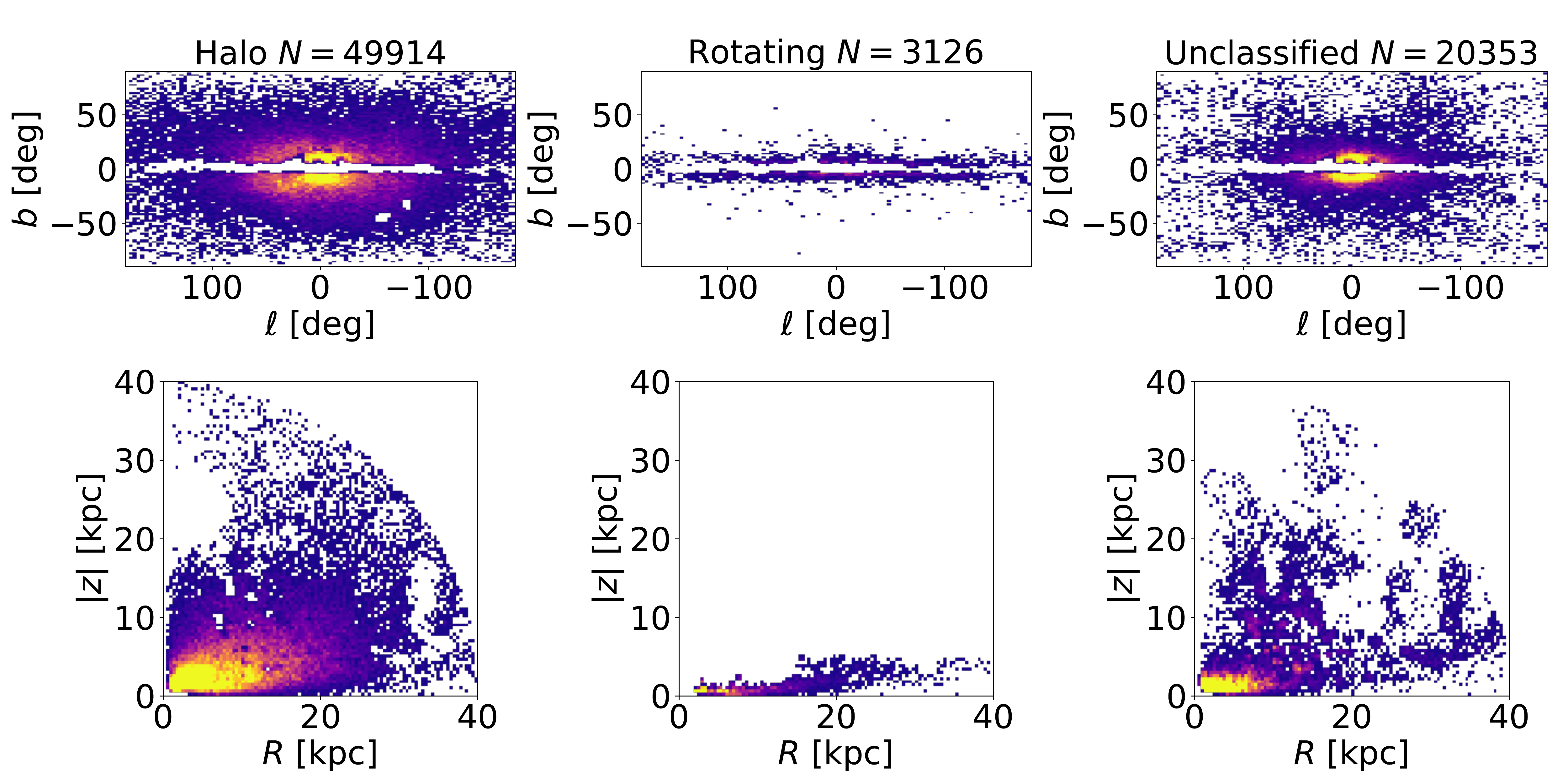}}
\caption[]{Three RRL groups. Same as Figure~\ref{fig:map_total} but
  for the stars in the Gclean catalogue (Section~\ref{sec:cleaning})
  belonging to the halo sub-sample (left), rotating disc-like
  subsample (centre) and stars that satisfy neither of the above
  criteria (right), see Section~\ref{sec:kseparation} for details. The
  color-map is the same as that shown in Figure~\ref{fig:map_total}.}
\label{fig:maps}
\end{figure*}

The final cleaned catalogue contains $72,973$ stars (Gclean
catalogue). We also produce a very conservative catalogue considering
only the stars that have been classified as RRab in both \gaia SOS and
$ASAS$-$SN$ ($17,570$ stars, SA catalogue), we also require that they
have complete \gaia lightcurve information (period and $\Phi_{31}$).
In the rest of the paper, we will compare the results of the analysis
of the two catalogues to investigate potential biases due to artefacts
and contaminants that went unnoticed.  The distributions of
heliocentric distances and of the transverse velocities in the Gclean
catalogue are shown in the bottom panel of
Figure~\ref{fig:hist_sample} (displaying the sample before the
distance cut).  Most of the stars are located within 20-25 kpc from
the Sun, but there are still hundreds of stars out to approximately 40
kpc; beyond this radius, the number of objects in the catalogue
decreases abruptly (these objects are not present in the final Gclean
catalogue). The relative distance and velocities uncertainties are
shown in the top panels of Figure~\ref{fig:hist_sample}: four
sequences are clear in the left-hand panel. The vertical sequence
located around 8-10 kpc is due to the stars in highly-extincted
regions where the uncertainties on the reddening dominate the error
budget (see Section~\ref{sec:dist_estimate}). The higher horizontal
sequence ($\delta D_\odot / D_\odot \approx 0.12$) comprises of the
stars without the period estimate. The other two sequences are due to
stars without $\Phi_\mathrm{31}$ estimate ($\delta D_\odot / D_\odot
\approx 0.11$) and to stars in the SOS catalogue with complete
information (period and $\Phi_\mathrm{31}$, $\delta D_\odot / D_\odot
\approx 0.10$).  Overall most of the stars have distance errors
slightly larger than 10\%, while the relative errors on velocities can
reach substantial values (up to $50-100\%$). The errors reported in
Figure~\ref{fig:hist_sample} are random errors based on the
Monte-Carlo analysis (Section~\ref{sec:dist_estimate}), however we
also analyse the possible systematic effects due to the assumptions
made when information about the period and/or when $\Phi_{31}$ and/or
the \gaia colors is not available
(Section~\ref{sec:dist_estimate}). For most of the cases, the
systematic shift is sub-dominant (relative error$\approx 5\%$) with
respect to the random errors. Hence, we do not include a systematic
component in the uncertainties used in the kinematic analysis. Based
on the error properties of the catalogue we expect that our analysis
(Section~\ref{sec:method}) is able to give reliable constraints on the
kinematic parameters within 20-30 kpc from the Galactic centre, while
the quality of the results progressively degrades at large radii.  The
distribution of the stars on the sky and in the Galactocentric $R,|z|$
plane are shown in the left-hand column of Figure~\ref{fig:map_total}.

\begin{figure}
\centering
\centerline{\includegraphics[width=1.0\columnwidth]{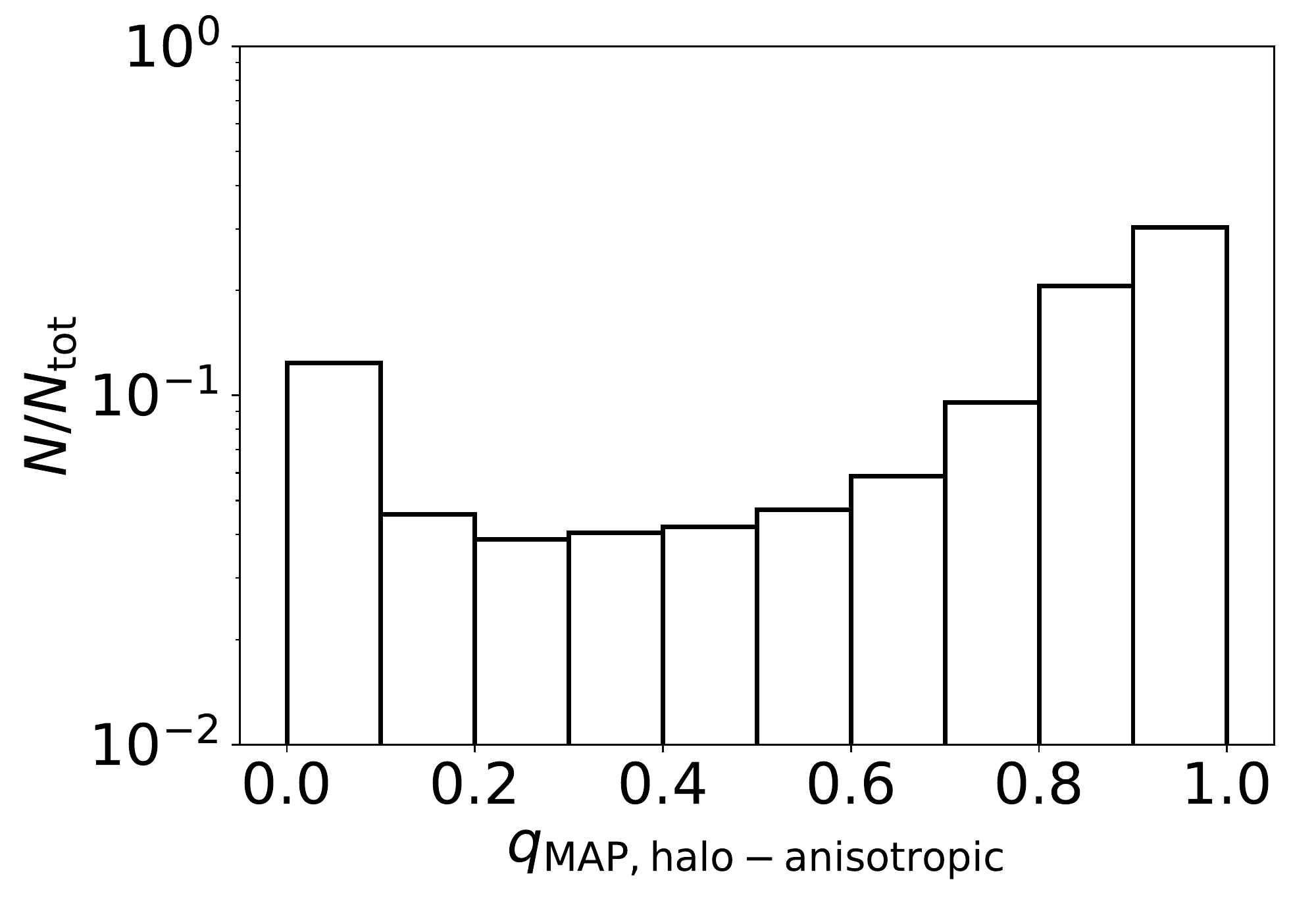}}
\caption[]{Distribution of the RRL maximum-a-posteriori probability
  (MAP, see Section~\ref{sec:kfit}) of belonging to the (radially)
  anisotropic kinematic component as inferred from the double
  component fit described in Section~\ref{sec:kseparation}. }
\label{fig:qhist_halo_radial}
\end{figure}
\begin{figure*}
\centering
\centerline{\includegraphics[width=0.9\textwidth]{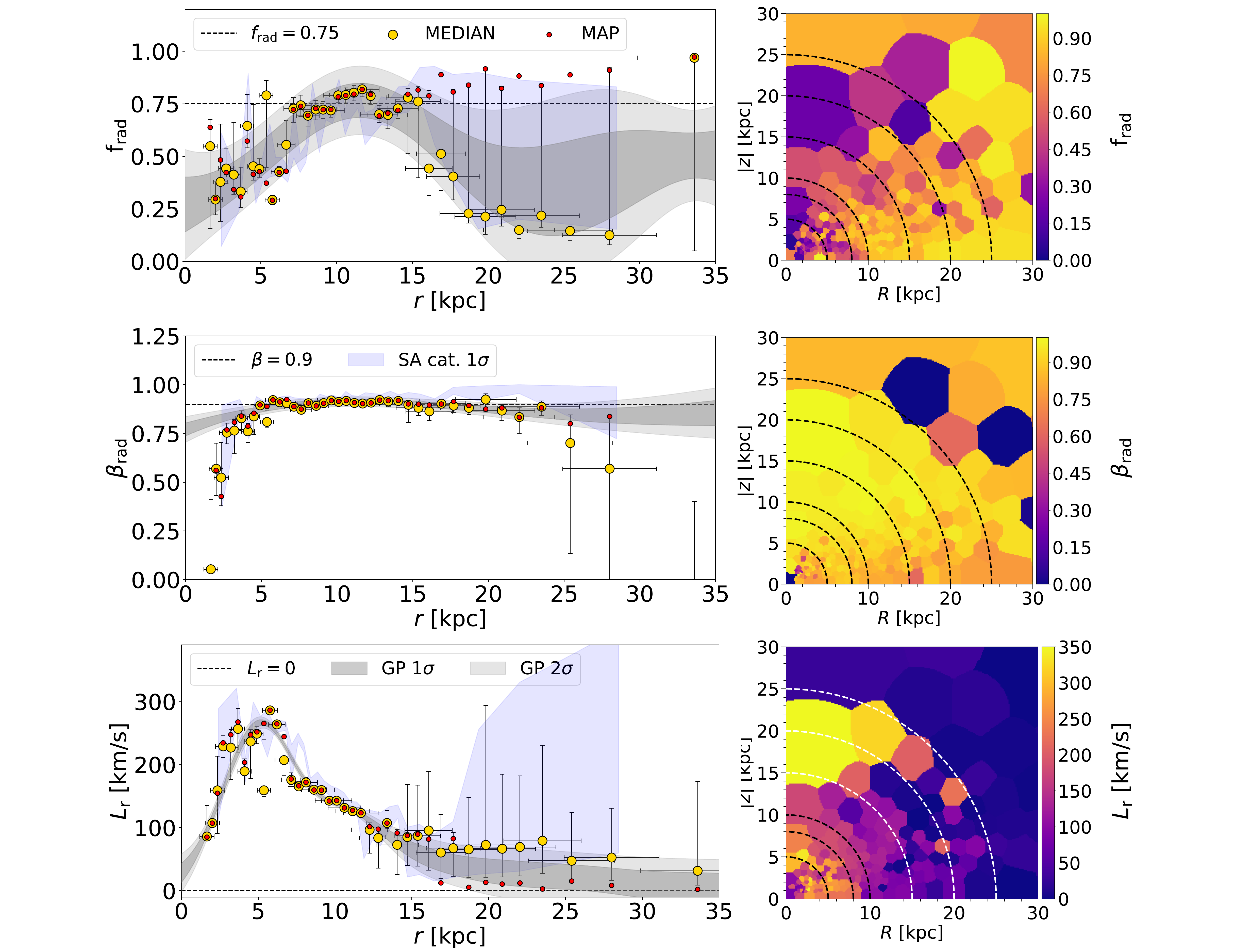}}
\caption[]{Properties of the radially-anisotropic halo component (see
  Section~\ref{sec:halo_kin}): relative fraction of the radial
  component over the total (top), its anisotropy (middle) and the
  position of the peak of the double-horn profile assumed for the
  distribution of the radial velocity (bottom, see Section~\ref{sec:halo}). Left (right) panels show the results of the model applied
  to spherical (cylindrical) Voronoi bins (see
  Section~\ref{sec:binning} and Section~\ref{sec:halo}). The large
  yellow data-points give the median of the a-posteriori distribution,
  while the error-bars indicate its 16th and 84th percentile; the
  small-red points show the Maximum-a-Posteriori (MAP) of the
  posteriors; X-axis represents the median of the spherical radial
  distribution, while the errorbars indicate the median value of the
  errors on the radius of the stars in each bin; the grey bands show
  the $1 \sigma$ and $2\sigma$ interval from a Gaussian Process (GP)
  interpolation. We interpolate the symmetrised version of the data
  points with a GP process: data-points show the middle values between
  the 16th and 84th percentile, while the vertical error-bars are half
  of the 16th-84th percentile distance; the blue band shows the
  1$\sigma$ interval of the posterior obtained using the SA
  (SOS+$ASAS$-$SN$) catalogue (see Section~ \ref{sec:cleaning}). The
  circular lines indicate the spherical radii of 5,8,10,15,20,25 kpc.}
\label{fig:halo_comb}
\end{figure*}

\section{The Method} \label{sec:method}

This work aims to study the kinematics of the RRL stars in the \gaia
dataset. Such an analysis is however hampered by the lack of
line-of-sight (los) velocity measurements for most of the stars in our
final catalogue -- indeed only 266 out of more than $70,000$ stars
have \gaia radial velocity. Relying on cross-matches with other
spectroscopic catalogue such as $RAVE$ \citep[][]{RAVE5}, $APOGEE$
\citep[][]{APOGEE}, or $LAMOST$ \citep[][]{LAMOST} would reduce the
number of objects as well as the radial extent and sky coverage of the
catalogue.  Moreover, the periodic radial expansion/contraction of the
RRL surface layers, if not taken into account, can bias the radial
velocity measurements by up to $40-70\ \kms$ \cite[see
  e.g.][]{Liu91,Drake2013}.


The lack of the los velocities makes it impossible to estimate the
full 3D velocity information on a star-by-star basis. However, since
stars at different celestial coordinates and different heliocentric
distances have distinct projections onto the 3D Galactic velocity
space, it is possible to estimate the velocity moments (mean values
and standard deviations) of the intrinsic 3D velocity ellipsoid using
the proper motions of a group of stars taken together under the
assumptions of symmetry \cite[see
  e.g.][]{Dehnen98,Schonrich12,Schonrich18,Wegg19}. In practice, we
consider two possibilities and assume that proper motions of stars i)
at the same $R$ and $|z|$ (cylindrical symmetry) or ii) the same $r$
(spherical symmetry) sample the same 3D velocity distribution.

\subsection{Kinematic fit} \label{sec:kfit}

In what follows we implement the ensemble velocity moment model
following and extending the method described in \cite{Wegg19} (\WG,
hereafter).  In this section we briefly summarise the method; further
details can be found in the original \WG paper.  The basic assumption
is that the intrinsic velocity distribution of stars in a given
Galactic volume at given Galactocentric coordinates (e.g.\ spherical
or cylindrical) is a multivariate normal
$f\left(\bm{V}\right)=\mathcal{N} \left( \bar{\bm{V}},\Sigma \right)$,
where $\bar{\bm{V}}$ is the Gaussian centroid and $\Sigma$ is the
covariance matrix or velocity dispersion tensor. This distribution can
be projected onto the heliocentric sky coordinates
$\bm{V}_\mathrm{sky}=(V_\mathrm{los}, V_\ell, V_b)$ appliyng the
rotation matrix $\mathbf{R}$ (different for each sky position)
satisfying $\bm{V}_\mathrm{sky}=\mathbf{R}\bm{V}$. The projected
distribution is still a Gaussian and therefore it can be easily
analytically marginalised over the unknown term
$V_\mathrm{los}$. Finally, the likelihood for a given star located at
given distance and position on the sky to have velocities
$\bm{V}_\mathrm{\perp}=(V_\ell, V_b)$ is given by
\begin{equation}
\mathcal{L} = \mathcal{N}\left( \bar{\bm{V}}_\mathrm{\perp}, \Lambda_\perp + \mathbf{S} \right),
\label{eq:lk}
\end{equation}
where 
\begin{itemize}
    \item $\bar{\bm{V}}_\mathrm{\perp} = \mathbf{R}_\perp
      \bar{\bm{V}}$ and $\mathbf{R}_\perp$ is the rotation matrix
      $\mathbf{R}$ without the 1st row related to the los velocity
      ($2\times3$ matrix, see Appendix
      \ref{appendix:rotation_matrix});
    \item $\Lambda_\perp$ is the projected covariance matrix  $\Lambda=\mathbf{R}\mathbf{\Sigma}\mathbf{R}^\intercal$ without the 1st row and the 1st column related to the los velocity  ($2\times2$ matrix);
    \item $\mathbf{S}$ is a 2x2 matrix of the $V_\ell, V_b$ measurement errors and covariance (see Section~ \ref{sec:dist_estimate}).
\end{itemize}
In order to estimate the velocity moments, we consider the total
likelihood as the product of the likelihoods (Equation~\ref{eq:lk}) of
all stars in a given Galactic volume bin. The method described so far
follows, point by point, what has been done in \WG. We add a further
generalisation considering the intrinsic velocity distribution as a
composition of multiple multivariate normal distributions. Therefore
the likelihood for a single star becomes
\begin{equation}
\mathcal{L}_\mathrm{multi} = \sum_i f_i \mathcal{N}\left( \bar{\bm{V}}_{i,\mathrm{\perp}}, \Lambda_{i,\perp} + \mathbf{S}_i \right)=\sum_i f_i \mathcal{L}_i,
\label{eq:lkmulti}
\end{equation}
where the component weights $f$ sum up to 1.  Using
Equation~\ref{eq:lkmulti} we can apply a Gaussian Mixture Model to the
intrinsic velocity distribution fitting only the observed tangential
velocities.  Starting form Equation~\ref{eq:lkmulti} it is possible to
define, for each star, the a-posterior likelihood of belonging to the
$i$th component as
\begin{equation}
q_i = \frac{f_i \mathcal{L}_i}{\mathcal{L}_\mathrm{multi}}. 
\label{eq:q}
\end{equation}
The stochastic variables $q$ (and their uncertainties) allow us to
decompose the stars into different kinematic populations using a
quantitative ``metric".  For a given sample of stars (see Section~
\ref{sec:binning}), we retrieve the properties ($\bm{V},
\mathbf{\Sigma}$) (3+6 parameters) of the kinematic components and
their weights adopting a Monte Carlo Markov Chain (MCMC) to sample the
posterior distributions generated by the product of all likelihoods
defined in Equation~\ref{eq:lkmulti}.  In practice, the posterior
distributions have been sampled using the affine-invariant ensemble
sampler MCMC method implemented in the \texttt{Python} module
\texttt{emcee}\footnote{\url{https://emcee.readthedocs.io/en/stable/}}
\citep{emcee}. We used 50 walkers evolved for 50000 steps after 5000
burn-in steps.  We evaluate the convergence of the chains by analysing
the trace plots and estimating the autocorrelation time
$\tau_f$\footnote{An useful note about autocorrelation analysis and
  convergence can be found at
  \url{https://emcee.readthedocs.io/en/stable/tutorials/autocorr/} }
(see e.g.\ \citealt{Goodman2010}). In particular, we check that for
all of our fits and parameters, the number of steps is larger than $50
\tau_f$, i.e. the number is sufficient to significantly reduce the
sampling variance of the MCMC run.  All kinematics models have been run
and analysed using the \texttt{Python} module
\texttt{Poe}\footnote{\url{https://gitlab.com/iogiul/poe.git}}.

In the next Sections, we exploit this method to separate the RRL
sample into two distinct kinematic components: a non-rotating (or
weakly rotating) halo-like population and a population with a large
azimuthal velocity. Subsequently, the same method is applied again to
separate kinematically the halo into an anisotropic and an isotropic
populations. The choice of binning in the given coordinate system
(spherical or cylindrical), the number of Gaussian components and the
prior distributions of their parameters are described in the following
Sections.

\begin{figure*}
\centering
\centerline{\includegraphics[width=1.0\textwidth]{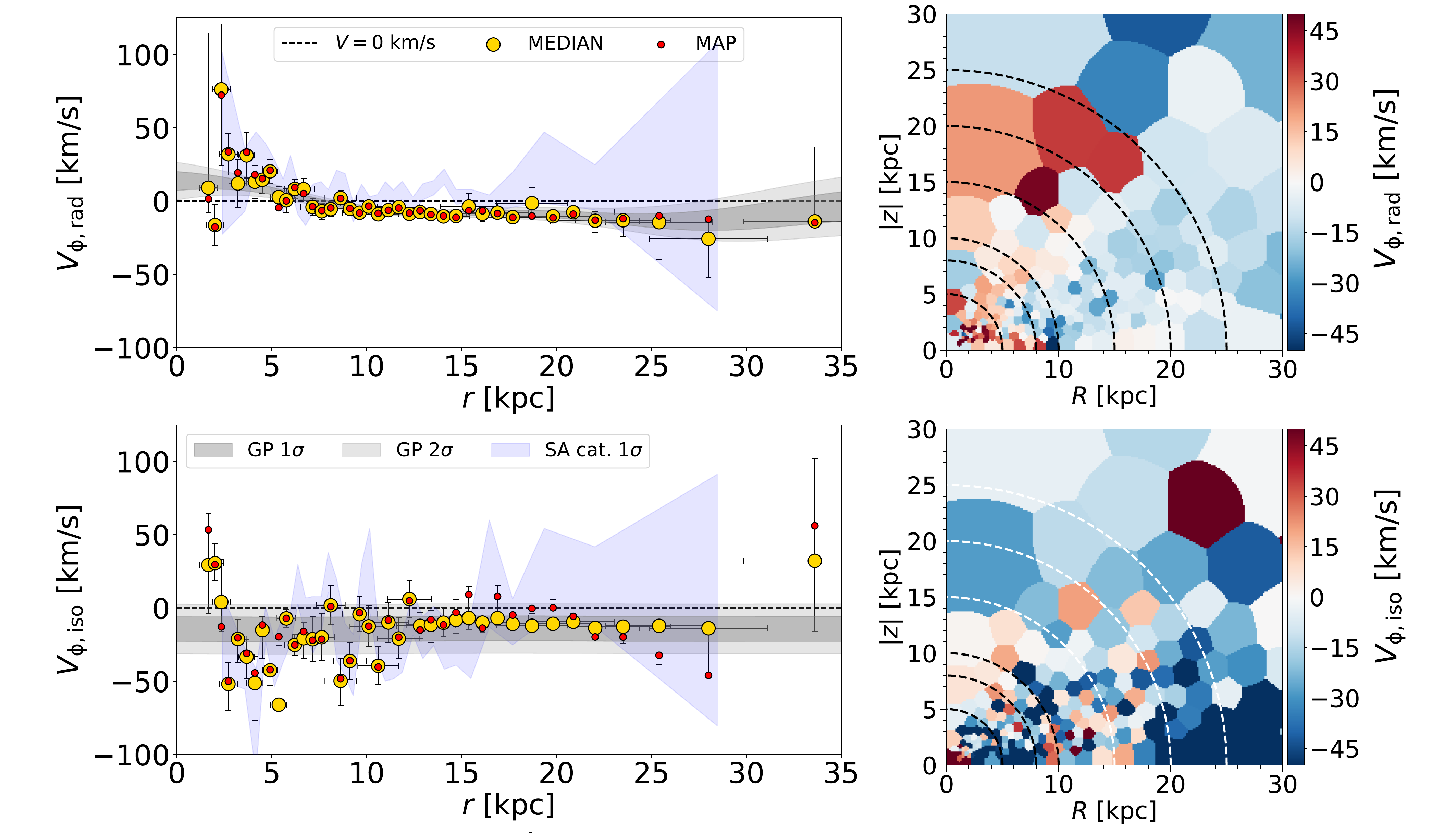}}
\caption[]{Same as Figure~\ref{fig:halo_comb} but for the azimuthal
  velocity for the radially-anisotropic (left-hand panel) and the
  isotropic (right-hand panel) components.}
\label{fig:halo_comb_vphi}
\end{figure*}

\subsection{Binning strategy} \label{sec:binning}

Each of our kinematic analyses is applied to stars grouped in bins of
Galactic $r$ or $R, |z|$ assuming spherical or cylindrical symmetry
correspondingly. In each of these bins the intrinsic distribution of
velocities is considered constant. In order to have approximately the
same Poisson signal-to-noise ratio ($\sqrt{N_\mathrm{stars}}$) in each
bin we compute a Voronoi tessellation of the $R, |z|$ plane making use
of the \texttt{vorbin} \texttt{Python} package
\citep{Cappellari03}\footnote{\url{https://www-astro.physics.ox.ac.uk/~mxc/software/\#binning}}. When
assigning stars to bins in spherical $r$, we select the bin edges so
that each bin contains $N_\mathrm{stars}$ objects. If the outermost
bin remains with a number of stars lower than $N_\mathrm{stars}$, we
merge it with the adjacent bin.  In the rest of the paper, we identify
the coordinates of a given bin ($R,|z|$ or $r$) as the median of the
coordinate of the stars in the bin, we associate to these values an
error that is the median of the corresponding errors of the
stars. Although we do not take account explicitly of the errors on
$R$,$|z|$ and $r$ in the kinematic fit, the velocities $V_\ell$ and
$V_b$ already incorporate the errors on distance
(Section~\ref{sec:dist_estimate}). In practice, we do not allow stars
to belong to more than one bin even if this is consistent with their
Galactic coordinate errors. This choice does not represent a serious
issue in our analysis, but at large radii, where the errors are
larger, the kinematic parameters obtained with our fit are likely
correlated in adjacent bins.

\subsection{Kinematic separation} \label{sec:kseparation}

\begin{table}
\centering
\begin{tabular}{lcc}
\hline
 & \multicolumn{2}{c}{Prior distributions} \\ \hline
\multicolumn{1}{l|}{} & \multicolumn{1}{c|}{halo} & \multicolumn{1}{l}{rotating} \\ \hline
\multicolumn{1}{l|}{$V_\mathrm{\phi}$} & \multicolumn{1}{c|}{$\delta(0)$} & $\mathcal{N}(100,200)[50,\infty]$ \\ \hline
\multicolumn{1}{l|}{$V_\mathrm{r}=V_\mathrm{\theta}$} & \multicolumn{2}{c}{$\delta(0)$} \\ \hline
\multicolumn{1}{l|}{$\sigma_\mathrm{r}$} & \multicolumn{1}{c|}{$\mathcal{N}(150,200)[0,\infty]$} & \multirow{2}{*}{$\mathcal{N}(0,20)[0,\infty]$} \\ \cline{1-2}
\multicolumn{1}{l|}{$\sigma_\mathrm{t}$} & \multicolumn{1}{c|}{$\mathcal{N}(100,200)[0,\infty]$} &  \\ \hline
\multicolumn{1}{l|}{$\rho_{\mathrm{r}\phi}=\rho_{\mathrm{r}\theta}=\rho_{\phi\theta}$} & \multicolumn{2}{c}{$\delta(0)$} \\ \hline
\multicolumn{1}{l|}{$f$} & \multicolumn{2}{c}{$\mathcal{U}(0,1)$} \\ \hline
\end{tabular}
\caption{Prior distributions for the parameters of the
  double-component fit: non-rotating halo/rotating components
  (Section~\ref{sec:kseparation}).  Both components are multivariate
  normals defined in a Galactocentric spherical frame of reference
  (see Section~\ref{sec:dist_estimate}). The parameters are from the top
  to the bottom: centroids of the normal distribution, velocity
  dispersions (assuming
  $\sigma_\mathrm{t}=\sigma_\mathrm{\phi}=\sigma_\mathrm{\theta}$ and
  $\sigma_\mathrm{r}=\sigma_\mathrm{t}$ for the isotropic component),
  covariance terms of the velocity dispersion tensor, weight of one of
  the component (see Equation~\ref{eq:lkmulti}). The used
  distributions are: Dirac Delta, $\delta$; normal,
  $\mathcal{N}(\bar{x},\sigma_{\mathrm{x}})$ where $\bar{x}$ is the
  centroid and $\sigma_\mathrm{x}$ the standard deviation; uniform,
  $\mathcal{U}(x_{\mathrm{low}},x_{\mathrm{up}})$ where
  $x_{\mathrm{low}}$ and $x_{\mathrm{up}}$ represent the distribution
  limits. The squared bracket indicate the distribution boundary,
  i.e. the prior probability is 0 outside the given range. If the
  brackets are not present the boundary is set to
  $[-\infty,\infty]$. All the velocity centroids and velocity
  dispersions are in unit of $\kms$.  Considering the parameters drawn
  from Dirac Delta as fixed in the fit, the total number of free
  parameters is 5.}
\label{tab:kseparation}
\end{table}

In order to separate the non-rotating halo from a component with a
high azimuthal velocity we set up a double-component fit:
\begin{itemize}
    \item 1st component (halo-like): spherical frame-of-reference, no
      rotation ($V_\mathrm{\phi}=0$), anisotropic velocity dispersion
      tensor (we fit the the radial, $\sigma_\mathrm{r}$, and
      tangential,
      $\sigma_\mathrm{t}=\sigma_\mathrm{\phi}=\sigma_\mathrm{\theta}$,
      velocity dispersion.).
    \item 2nd component (rotating): spherical frame of reference, isotropic velocity dispersion tensor. 
\end{itemize}
In both cases the centroids along $V_\mathrm{r}$ and
$V_\mathrm{\theta}$ are set to 0. We assume that the velocity
ellipsoids are aligned in spherical coordinates fixing to 0 the
diagonal terms of the velocity dispersion tensor \citep[see
  e.g.][]{EvansGaia}. Table~\ref{tab:kseparation} summarises the model
parameters and their prior distributions. In particular, we set
non-exchangeable priors for the velocity centroids and velocity
dispersions to break labelling degeneracy (switching between models in
the MCMC chains) and improve model identifiability\footnote{see
  \url{https://mc-stan.org/users/documentation/case-studies/identifying_mixture_models.html}
  for useful notes on identifiability of Bayesian Mixture Models.}. In
order to detect possible overfitting due to the double-component
assumption, we also run a single-component fit considering only the
halo model summarised in Table~\ref{tab:kseparation}.  The
significance of the more complex double component fit is analysed with
the Bayesian Information Criterion (BIC) using the
maximum-a-posteriori (MAP) of the likelihood,
$\mathcal{L}_\mathrm{MAP}$:
\begin{equation}
    \mathrm{BIC}=k \ln n - 2 \ln \mathcal{L}_\mathrm{MAP},
\end{equation}
where $k$ is the number of free parameters and $n$ is the data sample
size.  The model with the lowest BIC is preferred, in particular we
consider significant the results of the two component fit where the
BIC difference ($\Delta \mathrm{BIC}$) is larger than 10. In order to
apply the fit we separate the whole sample (72,973 stars) into 692
cylindrical $R, |z|$ bins with an average Poisson signal-to-noise
ratio of 10 (see Section~\ref{sec:binning}). The fit is applied
separately in each bin.

Figure~\ref{fig:disc_halo} presents the maps of the kinematic
properties of the two principal components, the halo and the disc in
cylindrical $R$ and $|z|$. The two rows give the same information, but
the bottom row shows the results of the double-component fit only if
there is a significant improvement as indicated by the Bayesian
Information Criterion $\Delta {\rm BIC}>10$, otherwise it reverts to
the results of a single-component fit. The first column shows the map
of the fractional contribution of the rotating component. While there
are some hints of rotating parts of the halo at high $|z|$ in the top
panel, as demonstrated by the bottom panel, these are not significant
enough. The bulk of the rotating component sits at $|z|<5$ kpc across
a wide range of $R$, and closer to the Sun its vertical extent is
clearly limited to a couple of kpc at most. The second column presents
the map of the azimuthal velocity $V_{\phi}$ as a function of $R$ and
$|z|$. Again, some Voronoi cells at high $|z|$ may have the kinematics
consistent with a slow rotation, however $\Delta{\rm BIC}$ criterion
renders them not significant enough. Therefore, in the bottom row,
these high $|z|$ cells are empty and the bulk of the $V_{\phi}$ map is
limited to low vertical heights where the rotation velocity is in
excess of $V_{\phi}>200$ kms$^{-1}$ across the entire range of
$R$. Two single bins at high $z$ with $R\approx10-15 \ \kpc$ survive
the BIC cut, they show an azimuthal rotation of $\approx 50
\ \kms$. Stars in these bins are likely related to the rotating halo
structure found in the unclassified sample and discussed in
Section~\ref{sec:unclass}.  Finally, the third column displays the
behaviour of the halo velocity anisotropy $\beta$ as mapped by
RRL. Except for a small region near the centre of the Milky Way and a
few cells at high $|z|$ where the motion appears nearly isotropic, the
rest of the halo exhibits strong radial anisotropy with
$0.6<\beta<0.9$.

Figure~\ref{fig:qhist_halo} shows the distribution of the posterior
probability of belonging to the non-rotating (halo) component for the
stars in our sample. 
{  Going from $q_\mathrm{MAP,halo}=1$ to $q_\mathrm{MAP,halo}=0$,
the distribution can be divided in three regions: a clear peak around $q_\mathrm{MAP,halo}=1$, these are the RRL that do not exhibit any significant rotation and thus can be confidently
assigned to the halo; a decreasing  trend in the number fraction ranging from $q_\mathrm{MAP,halo}\approx0.9$ to $q_\mathrm{MAP,halo}\approx0.5$; finally, a region with an increasing number fraction from $q_\mathrm{MAP,halo}\approx0.5$ to $q_\mathrm{MAP,halo}=0$. The latter region is likely populated by  the stars with disc-like
kinematics (closer to 0 is $q_\mathrm{MAP,halo}$, more robust is  the association with the rotating component), while the second region is composed of stars that do not fall squarely
into one of the two groups.  Setting this latter, undetermined group aside for now, we focus on the stars that can be classified as halo or disc with certainty. We select the halo and disc-like stars by applying the following cuts:
}


%
\begin{equation}
\begin{split}
\mathrm{halo} & : q_\mathrm{MAP,halo}>0.9 \ \& \ q_\mathrm{16th,halo} > 0.5 \\
\mathrm{rotating disc-like} & : q_\mathrm{MAP,halo}<0.3 \ \& \ q_\mathrm{84th,halo}<0.5  \\
\ \ & \&  \  \ |z|<5 \ \kpc
\ \& \ \Delta \mathrm{BIC}>10,
\end{split}
\label{eq:cuts}
\end{equation}
{ where $q_\mathrm{16th,halo}$ and $q_\mathrm{84th,halo}$ are the 16th and 84th percentile of the a-posteriori $q_\mathrm{halo}$ distribution.
The selection cut for the halo is 
straightforward (see Fig.\ 
\ref{fig:qhist_halo}), the additional cut on the 
16th percentile has been added to conservatively remove stars with poorly constrained 
$q_\mathrm{halo}$. The $q_\mathrm{MAP,halo}$ cut for the disc-like component   is somehow 
arbitrary but we find it  the  best compromise 
between a large enough number of stars (to have 
good statistics) and to be conservative enough 
to target the stars that are more ``purely" 
associated with the rotating component. The 
other conditions has been added to focus on the 
disc-like flattened structure ($z$ cut) and to 
remove portion of the Galaxy volume  where the presence of  two-component is not statistically significant  (BIC cut).}

Of the total 72,973 RRL in our sample, 49,914 (or $\approx68\%$) are
classified as halo, 3,126 (or $\approx4\%$) as disc; while the
remaining 19,993 ($\approx28\%$) are
unclassified. 
Figure~\ref{fig:maps} shows the distribution of the
three kinematic groups on the sky in Galactic coordinates (top row)
and in cylindrical $R, |z|$ (bottom row). The halo stars (first
column) span a wide range of Galactic latitudes but mostly reside in a
centrally concentrated, slightly flattened structure limited by $R<30$
kpc and $|z|<20$ kpc. 
{ The middle panels of Figure~\ref{fig:maps}
clearly show that the rotating component
has a disc-like spatial distribution and
extends to R $\approx$ 30 kpc (see also the bottom panels of Figure~\ref{fig:disc_halo})}.  Interestingly, a similarly-extended
and highly flattened distribution was already detected previously in
the sample of candidate-RRL stars selected in the first \gaia data
release \citep{Iorio18}.

Finally, the shape of the unclassified portion of our sample (third
column) resembles a superposition of the disc and the halo, albeit
more concentrated to the centre: most of the stars are at $R<10$ kpc
and $|z|<5$ kpc. Additionally, at higher $|z|$, there are several
lumps and lobes likely corresponding to parts of the Virgo
Overdensity and the Hercules Aquila Cloud
\citep[e.g.][]{Vivas2001,Vivas2006,HAC2007,Juric2008,Simion2014,Simion2019}.

\begin{figure*}
\centering
\centerline{\includegraphics[width=1.0\textwidth]{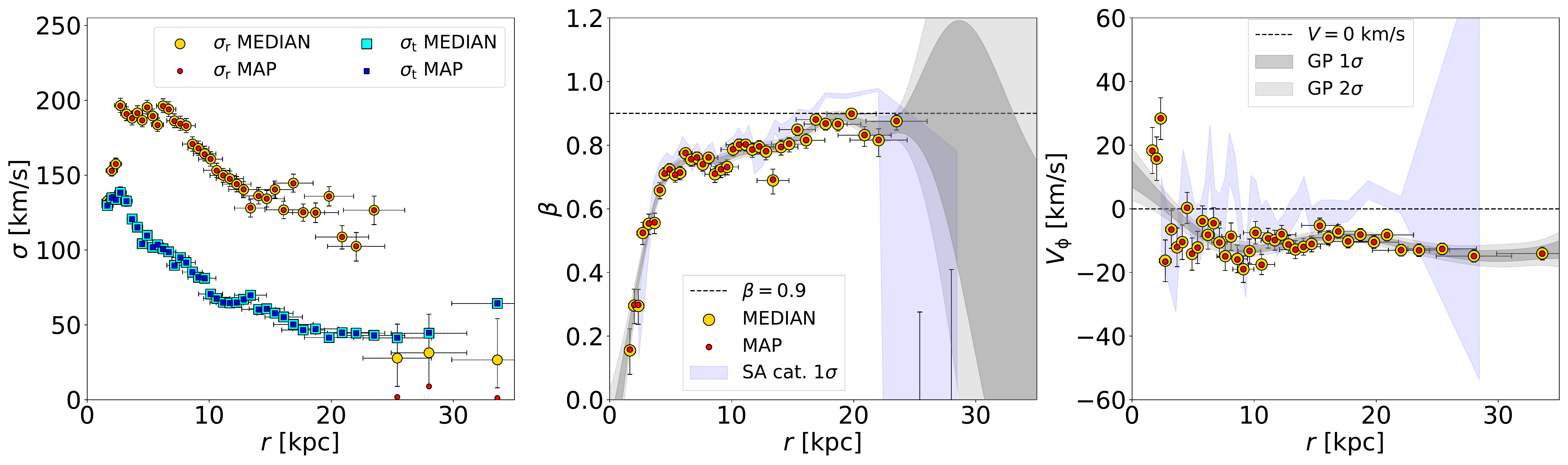}}
\caption[]{Same as Figure~\ref{fig:halo_comb} but for the anisotropy
  (middle panel) and the azimuthal velocity (right-hand panel)
  estimated in the single-component fit of the halo catalogue (see
  Section~ \ref{sec:halo}). The left-hand panel shows the radial and
  tangential velocity dispersion. }
\label{fig:halo_comb_1comp}
\end{figure*}

Our kinematic decomposition unambiguously demonstrates the presence of
a disc-like population amongst the \gaia RRL. According to the left
panel of Figure~\ref{fig:disc_halo}, this rapidly rotating population
contributes from $\approx30 \%$ (outer disc) to up to $\approx 50-60
\%$ (inner disc) of the RRL with $|z|<1$ kpc. We also see clear signs
of the RRL disc flaring beyond 15 kpc (see first two panels in the
bottom row of the Figure). This is unsurprising as the restoring force
weakens with distance from the Galactic centre \citep[see
  e.g.][]{Bacchini19}.  Additionally, the Milky Way disc at these
distances is withstanding periodic bombardment by the Sgr dwarf
\citep[e.g.][]{Laporte2018,Laporte2019}. The structure of the outer
disc as traced by RRL is consistent with the recent measurements of
the Galactic disc flare
\citep[e.g.][]{Lopez2014,Dekany19,Thomas2019,Skowron19}. In what
follows, we consider the halo and the disc RRL sub-samples, selected
using criteria listed in Equation~\ref{eq:cuts}, separately.

\section{The halo RR Lyrae} \label{sec:halo}

\begin{table}
\centering
\begin{tabular}{lcc}
\hline
 & \multicolumn{2}{c}{Prior distributions} \\ \hline
\multicolumn{1}{l|}{} & \multicolumn{1}{c|}{ halo-anisotropic} &  halo-isotropic \\ \hline
\multicolumn{1}{l|}{$V_\mathrm{\phi}$} & \multicolumn{1}{c|}{$\mathcal{N}(0,100)$} & $\mathcal{N}(0,100)$ \\ \hline
\multicolumn{1}{l|}{$V_\mathrm{r}=V_\mathrm{\theta}$} & \multicolumn{2}{c}{$\delta(0)$} \\ \hline
\multicolumn{1}{l|}{$L_\mathrm{r}\dagger$} & \multicolumn{1}{c|}{$\mathcal{N}(0,300)[0,\infty]$} & $\delta(0)$ \\ \hline
\multicolumn{1}{l|}{$\sigma_\mathrm{r}$} & \multicolumn{1}{c|}{$\mathcal{N}(150,100)[0,\infty]$} & \multirow{2}{*}{$\mathcal{N}(100,20)[0,\infty]$} \\ \cline{1-2}
\multicolumn{1}{l|}{$\sigma_\mathrm{t}$} & \multicolumn{1}{c|}{$\mathcal{N}(50,50)[0,\infty]$} &  \\ \hline
\multicolumn{1}{l|}{$\rho_{\mathrm{r}\phi}=\rho_{\mathrm{r}\theta}=\rho_{\phi\theta}$} & \multicolumn{2}{c}{$\delta(0)$} \\ \hline
\multicolumn{1}{l|}{$f$} & \multicolumn{2}{c}{$\mathcal{U}(0,1)$} \\ \hline
\end{tabular}
\caption{Same as Table~\ref{tab:kseparation} but for the double
  component fit: halo-anisotropic/halo-isotropic
  components. $\dagger$The halo-anisotropic component is a
  superposition of two multivariate normals (with same normalisation)
  offset from each other in $V_\mathrm{r}$ space by $2
  L_\mathrm{r}$ (see Section~\ref{sec:halo}). The total number of free parameters is 7.}
\label{tab:halo}
\end{table}

As convincingly demonstrated by \citet{Lancaster2019}, the kinematic
properties of the Galactic stellar halo can not be adequately
described with a single Gaussian. This is because the inner
$\approx30$ kpc are inundated with the debris from the {\it Gaia}
Sausage event \citep[see e.g.][]{Belokurov2018,SausageGCs}, also known
as {\it Gaia} Enceladus (see e.g. \citealt{Helmi2018,Koppelman20} but
see also \citealt{Evans2020}), producing a striking bimodal signature
in the radial velocity space. \citet{Lancaster2019} devise a flexible
kinematic model to faithfully reproduce the behaviour of an ensemble
of stars on nearly radial orbits \citep[see also][for a similar
  idea]{Necib2019}. We use the halo model developed by
\citet{Lancaster2019} and \citet{Necib2019} to describe the kinematics
of the halo sub-sample (see Section~\ref{sec:kseparation}). More
precisely, the model is the mixture of two components: isotropic and
anisotropic, both of which can rotate, i.e. have non-zero mean
$V_\mathrm{\phi}$.
The model, its parameters and their prior distributions are summarised
in Table~\ref{tab:halo}. The prior distributions of the anisotropic
component reflect our knowledge of the radially-anisotropic nature of
the halo. Moreover, they are set up to help the convergence of the
chain and the model identifiability as discussed in
Section~\ref{sec:kseparation}. By testing on the mock dataset we
ensure that the chosen priors are not preventing the selection of
isotropic ($\sigma_\mathrm{r}=\sigma_\mathrm{t}$) or
tangentially-anisotropic models
($\sigma_\mathrm{r}<\sigma_\mathrm{t}$) or models with simple Gaussian
distribution along $V_{\mathrm{r}}$ ($L_{\mathrm{r}}\approx0$).  This
two-component model with 7 free parameters is applied to the halo
sub-sample (49,914 stars) twice: once in bins of $r$ and again in bins
of $R$ and $|z|$ (see Section~\ref{sec:binning}). In the first case we
use 41 bins with an average Poisson signal-to-ratio of 35, in the
second case the bins are 203 with an average signal-to-ratio of
15. Parameters of both components are allowed to vary from bin to
bin. For comparison, we also model the RRL  kinematics in the halo
sub-sample with a single anisotropic multivariate normal with 4 free
parameters: $V_\mathrm{\phi}$ (prior $\mathcal{N}(0,100)$),
$\sigma_\mathrm{r}$, $\sigma_\mathrm{\phi}$, $\sigma_\mathrm{\theta}$
(prior $\mathcal{N}(0,200)[0,\infty]$).

{ Note that in our analysis, we do not attempt to distinguish between the bulge and the halo RR Lyrae.  This is because many of the classical bulge formation channels are not very different from those of the stellar halo, especially when both accreted and in-situ halo components are considered \citep[see e.g.][]{Kormendy2004,Athanassoula2005}. Historically, quite often the term ``bulge" is used to refer simply to the innermost region of the Milky Way. In that case, the Galactic bar and the discs would be included \citep[see e.g.][]{Barbuy2018}. However, we do not believe that these additional in-situ populations contribute significantly to the dataset we are working with. This is because our sample is highly depleted in the inner, low $|z|$ portion of the Galaxy where the RR Lyrae distribution is at its densest and the most complex, i.e. $R<2$ kpc. For example, we do not have any stars with $R<1$ kpc; there are only $\sim$2700 ($\sim$200) stars in the main (SA) sample with $R<2$ kpc.}

\subsection{Kinematic trends in the halo}\label{sec:halo_kin}

For stars in the halo sub-sample, Figure~\ref{fig:qhist_halo_radial}
shows the distribution of the posterior probability of membership in
either of the two components. As evidenced in the Figure, the
anisotropic component is dominant in this particular
dataset. Figure~\ref{fig:halo_comb} presents the properties of the
anisotropic halo population. Given the high values of $\beta$
displayed in the middle row of the Figure, we identify this component
with the \gaia Sausage debris \citep[see][for discussion of the GS as
  traced by the RRL]{Iorio19}.  It is important to note that, in some
cases, the median and the maximum-a-posteriori (MAP) points in
Figure~\ref{fig:halo_comb} show large differences because the
posterior distribution is bimodal. In those cases, the median results
are closer to the minimum that has been sampled more, while the
error-bars do not correspond to the classical Gaussian 1$\sigma$
errors but rather the distance between the two minima sampled by the
MCMC. Despite the large uncertainties due to the bimodal distribution,
the MAP and the median estimates indicate similar behaviour: if we
consider the MAP, the fraction of the radial component remains high
but $L_\mathrm{r}$ drops to 0; if we consider the median,
$L_\mathrm{r}\approx50 \ \kms$, but the fraction drops to small
values. Therefore, both the MAP and median indicate a transition
between the strong radially anisotropic component and the rest of the
stellar halo.
  
The top row of Figure ~\ref{fig:halo_comb} gives the contribution of
the stars in the radially-dominated portion of the halo as a function
of $r$. This fraction is at its lowest ($\approx20\%$) near the
Galactic centre. Outside of $R=3$ kpc, stars on nearly-radial orbits
contribute between $50\%$ and $80\%$. Beyond $R=20$ kpc, this fraction
becomes highly uncertain. From the right panel in the top row, it
appears that the contribution of the radially-biased debris falls
slightly faster with $|z|$, as expected if the debris cloud is
flattened vertically. The middle row of Figure~\ref{fig:halo_comb}
presents the behaviour of the velocity anisotropy $\beta$ with
Galactocentric radius $R$ (left) and $R$ and $|z|$ (right). Note that
in the model with two $V_\mathrm{r}$ humps, anisotropy $\beta$ can
increase i) when radial velocity dispersion dominates or ii) when the
velocity separation between the two humps $2L_r$ increases. For stars
in the radial component, $\beta$ is relatively low at
$\beta\approx0.3$ in the inner 3 kpc, but grows quickly to $\beta=0.9$
at 5 kpc and stays flat out to 20 kpc. Finally, the bottom panel of
the Figure shows the radial velocity separation $L_r$. It reaches
maximum $L_r\approx270$ kms$^{-1}$ around $3<R<5$ kpc from the
Galactic centre and then drops to $L_r\approx0$ kms$^{-1}$ around 30
kpc. The trend of $L_r$ as a function of $R$ looks very similar to the
projection of a high-eccentricity orbit onto the phase-space $(V_r,
R$). Along such an orbit, the highest radial velocity is reached just
before the pericentre crossing, where it quickly drops to zero. The
orbital radial velocity decreases more slowly towards the apocentre
where it also reaches zero. As judged by the bottom row of
Figure~\ref{fig:halo_comb}, the pericentre of the GS progenitor (in
its final stages of disruption) ought to be around $2<R<3$ kpc, while
its apocentre somewhere between $R=20$ kpc and $R=30$ kpc.
\begin{figure*}
\centering
\centerline{\includegraphics[width=1.0\textwidth]{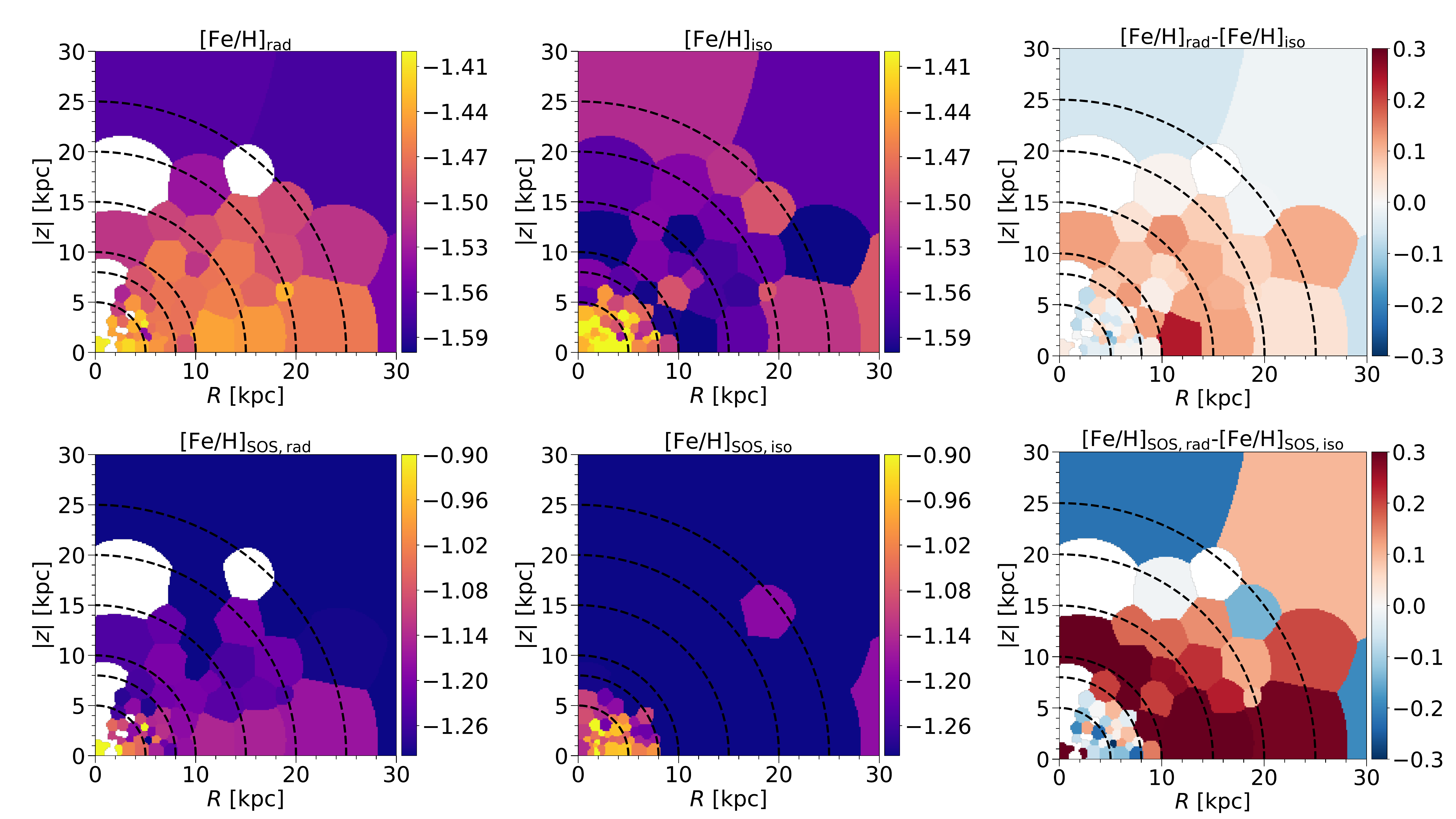}}
\caption[]{Cylindrical maps showing the distributions of the median
  metallicity estimated in this work (top, see Section~
  \ref{sec:dist_estimate} and Appendix \ref{appendix:met}) and
  reported in the SOS catalogue (bottom) respectively. Left-hand
  panels show the metallicity maps for the stars in the
  radially-biased halo component (23,734 stars) while the middle
  panels show the stars in the isotropic halo component (7,767
  stars). The right-hand panels show the difference between the radial
  and the isotropic component maps. The stars in this map are
  subsamples of the halo component (see Section~ \ref{sec:halo})
  belonging to the SOS catalogue and with an a-posteriori MAP
  likelihood of belonging to the anisotropic or isotropic component
  larger than 0.7 (see Figure~\ref{fig:qhist_halo_radial}). The
  Voronoi-tesselation has been obtained using the isotropic halo
  sample with a target Poisson signal-to-noise equals to 10. The bins
  in which the number of stars is lower than 50 are excluded from the
  maps (see e.g. the white bins in the left-hand and right-hand
  panels).}
\label{fig:halo_comb_met}
\end{figure*}

In Figure~\ref{fig:halo_comb} as well as in several subsequent Figures
we compare the kinematic properties of the \gaia DR2 RRL sample
(Gclean) with those obtained for a more restrictive set of RRL,
i.e. that produced by cross-matching the objects reported in the \gaia
SOS and by the $ASAS$-$SN$ variability survey (SA catalogue, shown as
light lilac filled contour). The SA catalogue does not only suffer
lower rate of contamination, it contains only bona fide RRab stars
with period information and, therefore, much more robust (and
unbiased) distance estimates. This more trustworthy RRL dataset comes
at a price: the size of the SA sample is $\approx5$ times smaller
compared to the Gclean catalogue and the sampled distances are reduced
by the magnitude limit ($V\approx17$) of the $ASAS$-$SN$ dataset.
Reassuringly, however, the differences between the kinematic
properties of the radially-biased halo component inferred with the
Gclean and the SA data are minimal as demonstrated in the left column
of Figure~\ref{fig:halo_comb}. The only clear distinction worth
mentioning is the blow-up of the $L_r$ confidence interval shown in
the bottom left panel. Beyond 15 kpc, the SA-based $L_r$ uncertainty
explodes due to the lack of distant RR Lyrae in this sample.

Figure~\ref{fig:halo_comb_vphi} is concerned with the mean azimuthal
velocity of each of the two halo components. Mean $V_\mathrm{\phi}$ is
shown for the radial (top) and the isotropic (bottom) portions of the
model applied to the halo sample. For the GS-dominated radially-biased
halo component, $V_\mathrm{\phi}$ is slightly prograde ($\approx15$
kms$^{-1}$) within the Solar circle and becomes slightly retrograde
($\approx-15$ kms$^{-1}$) outside of 10 kpc. Note that net rotation is
particularly affected by hidden distance biases \citep[as discussed in
  e.g.][]{AllegedDuality} and is driven by over- or under- correcting
for the Solar reflex motion (see Section~\ref{sec:tests}). The mean
azimuthal velocity of the radially-biased component of the halo plays
an important role in reconstructing the details of the GS merger. As
discussed in \citet{Belokurov2018}, the Sausage progenitor galaxy did
not necessarily have to arrive to the Milky Way head-on. Instead, the
dwarf could start the approach with plenty of angular momentum which
it then lost as it coalesced and disrupted in the Galaxy's
potential. The idea that dynamical friction could cause the orbit of a
massive satellite to {\it radialise} instead of {\it circularising}
was first proposed in \citet{Amorisco2017deposition}. A clearer
picture of the azimuthal velocity behavior is given by the SA dataset,
which is much less susceptible to distance errors, and as a
consequence to $V_{\phi}$ biases. The SA probability contours show
that the net rotation of the radially-biased halo component remains
very slightly prograde (at the level of $\approx15$ kms$^{-1}$)
throughout the Galactocentric distance range probed. Such slight
prograde spin is in agreement with a number of recent studies
\citep[see][]{SlightSpin,Tian2019,Wegg19,Belokurov2020}. Note that
this low-amplitude prograde rotation can only be claimed with some
degree of confidence at distances $R<10$ kpc, i.e. the region
containing a larger portion of RRL in our sample. Further out in the
halo, the net azimuthal velocity is consistent with zero \citep[see
  also][]{Bird2020,Naidu2020}. For the isotropic halo component, both
Gclean and SA datasets indicate a slight retrograde net rotation
($\approx-20$ kms$^{-1}$), at least in the inner Galaxy.

Figure~\ref{fig:halo_comb_1comp} offers a view of the Galactic stellar
halo as described by a single Gaussian component\footnote{The fit
  parameters and their prior distributions are the same of the
  anisotropic halo component summarised in Table~ \ref{tab:halo} but
  with $L_\mathrm{r}\sim\delta(0)$. The total number of free parameter
  is 3.}.  It is not surprising to see the behaviour which appears to
be consistent with an average between the strongly radial and
isotropic components shown in the previous Figures. Between 5 and 25
kpc, the velocity anisotropy is high $0.75<\beta<0.9$, only slightly
lower than that shown in the top left panel of
Figure~\ref{fig:halo_comb_vphi}. Similarly, the superposition of
slightly prograde and slightly retrograde populations yields a mean
azimuthal velocity consistent with zero \citep[as previously reported
  e.g. by][]{Smith2009} as measured for the SA sample (see filled pale
lilac contours in the right panel of the Figure). The Gclean dataset
gives a retrograde bias of $-10$ kms$^{-1}$. Remember however that a
portion of the halo was excised and is now a part of the
`unclassified' subset. These `unclassified' RRL ought to be considered
to give the final answer as to the net rotation of the halo (see
Section~\ref{sec:unclass}).

\begin{figure*}
\centering
\centerline{\includegraphics[width=1.0\textwidth]{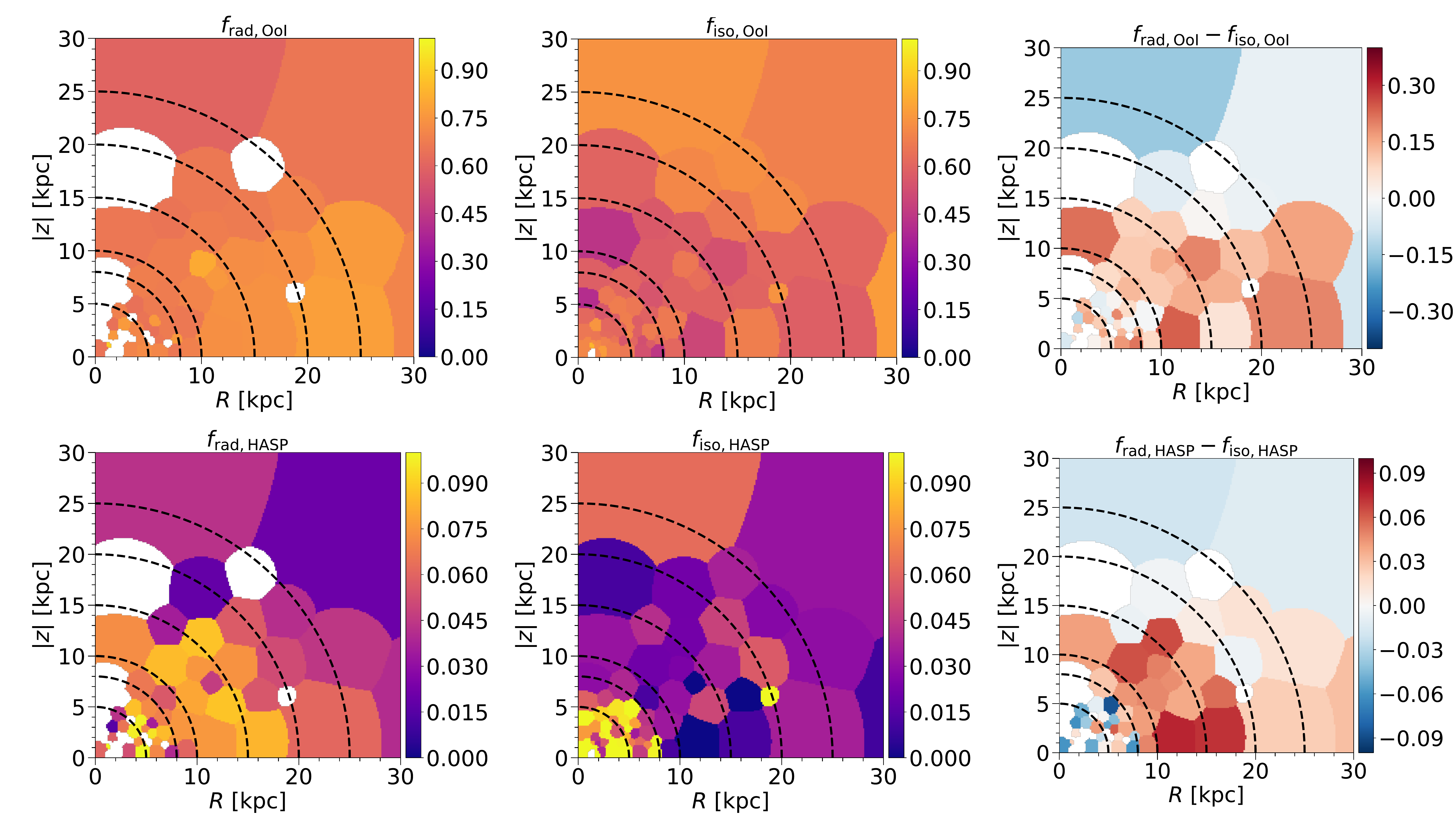}}
\caption[]{Same as Figure~\ref{fig:halo_comb_met} but for the
  Oosterhoff Type 1 (OoI, top panels) and the High Amplitude Short
  Period (HASP, bottom panels) fractions. See
  Section~\ref{sec:halo_pop} and \cite{BelokurovHASP}.}
\label{fig:halo_comb_fHASP}
\end{figure*}

\subsection{Stellar population trends in the halo} \label{sec:halo_pop}

\citet{Belokurov2018} used $SDSS$+\gaia DR1 data to establish a tight
link between the velocity anisotropy and the metallicity in the local
stellar halo. They show that the highest values of $\beta\approx0.9$
are achieved by stars with metallicity $-1.7<$[Fe/H]$<-1.2$, while at
lower metallicities the anisotropy drops to $0.2<\beta<0.4$. Using a
suite of zoom-in simulations of the MW halo formation, the prevalence
in the Solar neighborhood of comparatively metal-rich halo stars on
highly eccentric orbits is interpreted by \citet{Belokurov2018} as
evidence for an ancient head-on collision with a relatively massive
dwarf galaxy. In this picture, the lower-anisotropy and
lower-metallicity halo component is contributed via the accretion of
multiple smaller Galactic sub-systems. Note that strong trends between
orbital and chemical properties in the Galactic stellar halo had been
detected well before the arrival of the \gaia data \citep[see
  e.g.][]{ELS1962,Chiba2000,Ivezic2008,Bond2010,Carollo2010}. Most
recently such chemo-kinematic correlations have been observed in
glorious detail in multiple studies that used the GDR2 astrometry
\citep[e.g.][]{HaloAction,Deason2018,Lancaster2019,Conroy2019,Das2020,Bird2020,Feuillet2020}. Consequently,
in the last couple of years, a consensus has emerged, based on the
numerical simulations of stellar halo formation and chemical evolution
models, that the bulk of the local stellar halo debris is contributed
by a single, old and massive (and therefore relatively metal-rich)
merger
\citep[see][]{Haywood2018,Helmi2018,Mackereth2019,Fattahi2019,Bignone2019,Bonaca2020,Renaud2020,Elias2020,Grand2020}.

Figure~\ref{fig:halo_comb_met} explores the connection between the RR
Lyrae kinematics and their metallicity (estimated from the lightcurve
shape, see Section~ \ref{sec:dist_estimate} and Appendix
\ref{appendix:met}).  Both the top and the bottom row use the sample
of halo stars contained in the SOS catalogue of \gaia DR2 RRL. In the
top row, we present the metallicity maps obtained using our [Fe/H]
calibration presented in Equations ~\ref{eq:Met_RRab} and
~\ref{eq:Met_c}. The bottom row uses the metallicity estimates
reported as part of the SOS catalogue. While the two rows display
different absolute mean values of [Fe/H] in the halo (due to different
calibrations used), the relative metallicity changes as a function of
$R$ and $|z|$ and between the two halo components look very
similar. The left column of Figure~\ref{fig:halo_comb_met} shows the
metallicity distribution in the radially-biased halo component. As
discussed above, the bulk of this halo population has likely been
contributed by the \gaia Sausage merger. Both top and bottom panels
reveal a slightly flattened ellipsoidal structure whose metallicity is
elevated compared to the rest of the halo. This [Fe/H] pattern extends
out to $R\approx30$ kpc and $|z|\approx20$ kpc. No significant
metallicity gradient is observed in the radial direction, although the
inner 2-3 kpc do appear to be more metal-rich. However, given the
behaviour of $L_r$ shown in Figure~\ref{fig:halo_comb}, we conjecture
that very little \gaia Sausage debris reaches the inner core of the
Galaxy (see Section~\ref{sec:halo_kin} for discussion). In the
vertical direction, there are hints of a metallicity gradient where
[Fe/H] decreases with increasing $|z|$.

The behaviour of [Fe/H] in the isotropic halo component is given in
the middle column of Figure~\ref{fig:halo_comb_met}. The most striking
feature in the metallicity distribution of the isotropic component is
the compact spheroidal structure with $R<10$ kpc whose mean
metallicity exceeds that of the radially-anisotropic component (and
hence that of the \gaia Sausage).  Beyond $R\approx10$ kpc, no strong
large-scale metallicity gradient is discernible: [Fe/H] does change
appreciably and stays at levels slightly lower than those achieved by
the GS debris at similar spatial coordinates. To contrast the
metallicity trends of the two halo components, the right column of the
Figure shows the difference of the left and middle metallicity
distributions. This differential picture highlights dramatically the
shape of the GS debris cloud whose mean metallicity sits some $0.2$
dex above the typical halo [Fe/H] value. Even more metal-rich is the
inner 10 kpc. This inner halo structure - which also appears flattened
in the vertical direction - exhibits the highest mean metallicity in
the inner 30 kpc of the halo, at least $0.2$ dex higher than the
radially-biased GS.

The position of an RRL  on the period-amplitude plane contains
non-trivial information about its birth environment. In the Milky Way
halo, globular clusters show a well-defined `Oosterhoff dichotomy' 
\citep[][]{Oo1939,Oo1944} where RRL  in clusters of Oosterhoff
Type I (OoI) have a shorter mean period compared to those in GCs of
Oosterhoff Type II (OoII). The `Oosterhoff dichotomy' is not present
in the dwarf spheroidals observed today around the Milky Way that
appear to contain mixtures of Oosterhoff types but not in arbitrary
proportions \citep[e.g.][]{Catelan2004,Catelan2009}. Thus, the
relative fraction of RRL  of each Oosterhoff type can be used to
decipher the contribution of disrupted satellite systems to the
Galactic stellar halo \citep[see e.g.][]{Miceli2008,
  Zinn2014}. Finally, the so-called High Amplitude Short Period (HASP)
RRL  can be found across the Milky Way but are rather rare amongst
its satellites. This allowed \citet{Stetson2014} and
\citet{Fiorentino2015} to put constraints on the contribution of dwarf
galaxies of different masses to the Galactic stellar halo. Most
recently, \citet{BelokurovHASP} used RRL  tagging according to
their type (OoI, OoII or HASP) to `unmix' the Milky Way halo. Taking
advantage of the wide-area RRL  catalogue provided as part of the
Catalina Real-Time Transient Survey
\citep[][]{Drake2013,Drake2014,css}, they show that the fraction of
OoI RRL  changes coherently and dramatically as a function of
Galactocentric distance. They also demonstrate that in the Milky Way
dwarf spheroidal satellites, the OoI fraction increases with dwarf's
mass. Using a suite of Cosmological zoom-in simulations,
\citet{BelokurovHASP} conjecture that the radial evolution in the RR
Lyrae mixture is driven by a change in the fractional contribution of
satellites of different masses. More precisely, they interpret the
peak in the OoI fraction within $R\approx30$ kpc as evidence that the
Milky Way's inner halo is dominated by the debris of a single massive
galaxy accreted some 8-11 Gyr ago. This picture is confirmed by the
change in the HASP RRL  at $10<R$(kpc)$<30$. However, inwards of
$R\approx10$ kpc, the HASP fraction grows further, to levels
significantly higher than those displayed in the most massive MW
satellites such as LMC, SMC and Sgr, making the very core of the halo
unlike any satellite on orbit around the Galaxy today. { Note that the Oosterhoff and HASP classes are used here simply as a way to select particular regions on the period-amplitude plane. The exact position on this so called Bailey diagram has remained a useful RR Lyrae diagnostic tool for decades but is only now starting to be investigated thoroughly with the help of the {\it Gaia} data and high-resolution spectroscopy \citep[see e.g.][]{Fabrizio2019}.}

Figure~\ref{fig:halo_comb_fHASP} follows the ideas discussed in
\citet{BelokurovHASP} and tracks the fraction of OoI type (top) and
HASP (bottom) RRL  as a function of $R$ and $|z|$ in both
radially-biased (left) and isotropic (middle) halo
components. Additionally, the difference between the two maps is shown
in the right column of the Figure. As the Figure demonstrates, the OoI
and HASP fractions in the radially-biased halo component are higher
compared to the isotropic halo population. In comparison, the RRL 
in the inner $\approx10$ kpc show slightly lower OoI contribution, yet
the HASP fraction is higher. These trends in the
period-amplitude of halo RRL  are fully consistent with those
presented in \citet{BelokurovHASP} and support the picture in which
the RRL  on highly eccentric orbits originate from a single
massive and relatively metal-rich dwarf galaxy. Given its lower
metallicity, lower fraction of OoI and HASP RRL,  the isotropic
population could be a superposition of tidal debris from multiple
smaller sub-systems.

\begin{figure*}
\centering
\centerline{\includegraphics[width=1.0\textwidth]{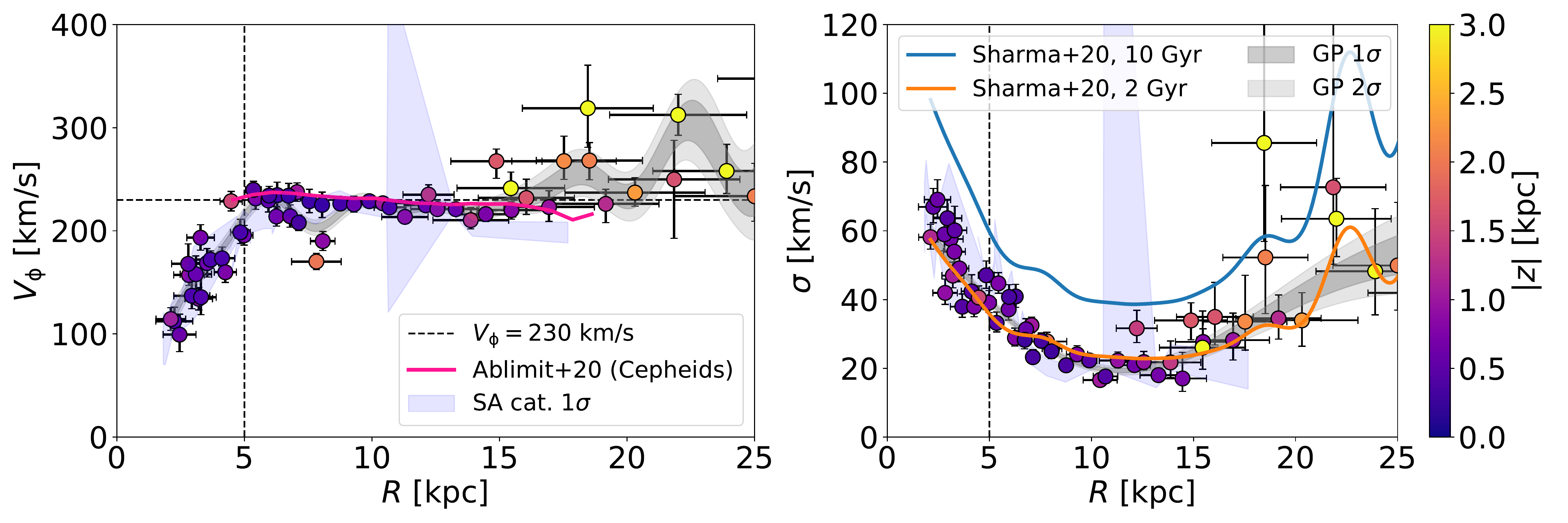}}
\caption[]{Azimuthal velocity and velocity dispersion (assuming
  isotropy) obtained for the sample of rotating stars (see
  Section~\ref{sec:disc}). Y-axis gives the median of the a-posteriori
  distribution of the azimuthal velocity, while the errorbars indicate
  its 16th and 84th percentile.  X-axis shows the median of the
  cylindrical radial distribution, while the error-bars indicate the
  median value of the errors on the cylindrical radius of the stars in
  the given bin.  Vertical black dashed lines mark 5 kpc radius
  roughly corresponding to the region where the presence of the bar
  may be important. The horizontal dashed line in the left panel
  indicates $V_\mathrm{\phi}=230 \ \kms$. Grey bands show the $1
  \sigma$ and $2\sigma$ intervals from the Gaussian Process
  interpolation as described in Figure~\ref{fig:halo_comb}. Blue band
  shows the 1$\sigma$ interval of the posterior obtained using the SA
  (SOS+$ASSASN$) catalogue (see text). The blue SA band explodes
  around $R\approx12$ kpc due to a particular bin where most of stars
  have been classified as the background. The magenta line in the
  left-hand panel shows the azimuthal velocity measured by
  \cite{Ablimit20} using a sample of Cepheids. The blue and orange
  lines in the right-hand panel show the median of the combination of
  the vertical and radial velocity dispersion model predictions by
  \cite{Sharma20}. Here we assumed [Fe/H]=$-1.0$ (see
  Figure~\ref{fig:disc_halo_compare}), $z=0.5\ \kpc$ and stellar age
  $t=2 \ \Gyr$ (orange line) and $t=10 \ \Gyr$ (blue line), see text
  in Section~\ref{sec:disc} for further information.}
\label{fig:disc_vrot}
\end{figure*}

As Figures~\ref{fig:halo_comb},~\ref{fig:halo_comb_met}
and~\ref{fig:halo_comb_fHASP} reveal, the inner $5$-$10$ kpc of the
Galactic stellar halo look starkly distinct from both the metal-richer
radially-biased \gaia Sausage debris cloud and the metal-poorer
isotropic halo. \citet{BelokurovHASP} suggested that a third kind of
accretion event is required to explain the RRL properties in the inner
Milky Way. This hypothesis, however, must be revisited in light of the
\gaia data.  Thanks to the \gaia DR1 and DR2 astrometry, we now have a
better understanding of the composition of the Galactic stellar halo
within the Solar radius. In particular, there now exist several lines
of evidence that perhaps as much as $\approx50\%$ of the nearby halo
could be formed in situ. The earliest evidence for such a dichotomy in
the stellar halo could be found in \citet{Nissen2010} who identified
two distinct halo sequences in the $\alpha$-[Fe/H] abundance
plane. Using \gaia DR1 astrometry complemented with $APOGEE$ and
$RAVE$ spectroscopy, \citet{Bonaca2017} showed that approximately half
of the stars on halo-like orbits passing through the Solar
neighborhood are more metal-rich than [Fe/H]$=-1$ and were likely born
in-situ. \citet{Babusiaux18} used \gaia DR2 data to build a
colour-magnitude diagram of nearby stars with high tangential
velocities and showed that the Main Sequence of the
kinematically-selected halo population is strongly
bimodal. Subsequently, \citet{Haywood2018}, \citet{DiMatteo2019} and
\citet{Gallart2019} used \gaia DR2 to investigate the behaviour of the
stars residing in the blue and red halo sequences uncovered by
\citet{Babusiaux18}. All three studies agreed that the blue sequence
is provided by the accreted tidal debris while the stars in the red
sequence were likely formed in-situ. Both \citet{DiMatteo2019} and
\citet{Gallart2019} point out that the stars in the in-situ component
had likely formed before the accretion of \gaia Sausage and were
heated up onto halo orbits as a result of the merger. It remains
somewhat unclear however where the thick disc stops and the in-situ
halo starts.

\citet{Belokurov2020} used the catalogue of stellar orbital properties
and accurate ages produced by \citet{Sanders2018} to isolate the halo
component they dubbed the `Splash'. Splash contains stars with high
metallicities $-0.7<[Fe/H]<-0.2$ and low-angular momentum (or
retrograde) motion. Importantly, its azimuthal velocity distribution
does not appear to be an extension of the thick disc's -- it stands
out as a distinct kinematic component \citep[see
  also][]{Amarante2020}. The age distribution of the Splash population
shows a sharp drop around 9.5 Gyr in agreement with previous estimates
described above. \citet{Belokurov2020} used Auriga
\citep[][]{Grand2017} and Latte \citep[][]{Wetzel2016} numerical
simulations of Milky Way-like galaxy formation to gain further insight
into the Splash formation. They demonstrate that a Splash-like
population is ubiquitous in both simulation suites and indeed
corresponds to the ancient Milky Way disc stars `splashed' up onto the
halo-like orbits \citep[as conjectured by
  e.g.][]{Bonaca2017,DiMatteo2019, Gallart2019}. Most recently,
\citet{Grand2020} provided a detailed study of the effects of the
\gaia Sausage-like accretion events on the nascent Milky Way. They
show that the propensity to Splash formation can be used to place
constraints on the properties of the \gaia Sausage accretion event,
for example the mass ratio of the satellite and the
host. Additionally, they demonstrate that in many instances in their
suite, the accretion is gas-rich and leads to a star-burst event in
the central Milky Way. Interestingly, as pointed out by
\citet{Belokurov2020}, recent observations of intermediate-redshift
galaxies reveal that star-formation can originate in the gas outflows
associated with profuse AGN or star-formation activity
\citep[see][]{Maiolino2017,Gallagher2019,Veilleux2020} thus raising a
question of whether the Milky Way's Splash could also originate in the
gas outflow \citep[see also][]{Yu2020}.

While the earlier studies of the Galactic in-situ halo had been
limited to the Solar neighborhood
\citep{Nissen2010,Bonaca2017,Haywood2018, DiMatteo2019,Gallart2019},
\citet{Belokurov2020} provide the first analysis of the overall
spatial extent of this structure. Using a selection of spectroscopic
datasets, they show that the Splash does not extend much beyond
$R\approx15$ kpc and $|z|\approx10$ kpc. Compare the picture in which
the Splash looks like a miniature halo - or perhaps a blown-up bulge -
(see red contours in Figures 11 and 13 in \citealt{Belokurov2020}) and
the RRL  stellar population maps presented here in
Figures~\ref{fig:halo_comb_met} and ~\ref{fig:halo_comb_fHASP}. There
is a very clear correspondence between the metal-rich and
HASP-enhanced portion of the (mostly) isotropic halo population and
the Splash. We therefore conjecture that the inner $\approx$10 kpc of
the Galactic halo RRL  distribution is pervaded by the in-situ
halo population. The in-situ halo RRL  are metal-rich and have
lower mean OoI fraction compared to \gaia Sausage and possess the
highest mean HASP fraction amongst all halo components.

\section{The disc RR Lyrae} \label{sec:disc}

\begin{table}
\centering
\begin{tabular}{lcc}
\hline
 & \multicolumn{2}{c}{Prior distributions} \\ \hline
\multicolumn{1}{l|}{} & \multicolumn{1}{c|}{disc} & background \\ \hline
\multicolumn{1}{l|}{$V_\mathrm{\phi}$} & \multicolumn{1}{c|}{$\mathcal{N}(0,400)[0,\infty]$} &  \\ \hline
\multicolumn{1}{l|}{$V_\mathrm{R}=V_\mathrm{z}$} & \multicolumn{1}{c|}{$\delta(0)$} &  \\ \hline
\multicolumn{1}{l|}{$\sigma=\sigma_\mathrm{R}=\sigma_\mathrm{z}=\sigma_\mathrm{\phi}$} & \multicolumn{1}{c|}{$\mathcal{N}(0,200)[0,\infty]$} &  \\ \hline
\multicolumn{1}{l|}{$\rho_{\mathrm{R}\mathrm{z}}=\rho_{\mathrm{R}\phi}=\rho_{\mathrm{z}\phi}$} & \multicolumn{1}{c|}{$\delta(0)$} &  \\ \hline
\multicolumn{1}{l|}{$\bar{V_\ell}$} & \multicolumn{1}{c|}{} & $\delta(\langle V_{\ell,\mathrm{stars}} \rangle)$ \\ \hline
\multicolumn{1}{l|}{$\bar{V_b}$} & \multicolumn{1}{c|}{} & \multicolumn{1}{l}{$\delta(\langle V_{b,\mathrm{stars}} \rangle)$} \\ \hline
\multicolumn{1}{l|}{$\sigma_\ell$} & \multicolumn{1}{c|}{} & $\mathcal{C}(0,500)[0,\infty]$ \\ \hline
\multicolumn{1}{l|}{$\sigma_b$} & \multicolumn{1}{c|}{} & $\mathcal{C}(0,500)[0,\infty]$ \\ \hline
\multicolumn{1}{l|}{$\rho_{\ell,b}$} & \multicolumn{1}{c|}{} & $\mathcal{U}(-1,1)$ \\ \hline
\multicolumn{1}{l|}{$f$} & \multicolumn{2}{c}{$\mathcal{U}(0,1)$} \\ \hline
\end{tabular}
\caption{ Same as Table~\ref{tab:kseparation} but for the parameters
  of the double component fit: rotating-disc/background. The
  rotating-disc component is a 3D multivariate normal distribution
  defined in a Galactocentric cylindrical frame of reference (see
  Section~ \ref{sec:dist_estimate}) with parameters: centroids
  ($V_\mathrm{\phi}$,$V_\mathrm{R}$,$V_\mathrm{z}$), isotropic
  velocity dispersion $\sigma$ and correlation terms of the velocity
  dispersion tensor $\rho$. The background is modelled as 2D
  multivariate normal in the observed velocity space. The parameters
  are the centroids ($\bar{V_\ell},\bar{V_b}$), which are fixed to the
  average values of the observed velocity distribution of the stars in
  each bin, the velocity dispersions ($\sigma_\ell,\sigma_b$) and the
  velocity correlation ($\rho_{\ell,b}$).
  $\mathcal{C}(x_\mathrm{c},l)$ indicates the Chaucy distribution
  centred in $x_\mathrm{c}$ and with scale $l$.  The total number of
  free parameters is 6. }
  \label{tab:prior_disc}
\end{table}

As described in Section~\ref{sec:kseparation}, a small but significant
fraction of the GDR2 RRL  (just under $5\%$) are classified as
belonging to a rotating component based on their
kinematics. Figures~\ref{fig:disc_halo} and \ref{fig:maps}  demonstrate
that the stars in the rotating sample are heavily biased towards low
Galactic latitude $|b|$ and small height $|z|$ and thus likely
represent a Milky Way disc population. Here we provide a detailed
discussion of the properties of this intriguing specimen.

In order to take into account possibile residual contaminants and
outliers in the sample of rotating RRL (see Section~
\ref{sec:kseparation}) we set a double component fit (see
e.g.\ \citealt{Hogg10}):
\begin{itemize}
    \item 1st component (disc-like): cylindrical frame-of-reference,
      isotropic velocity dispersion tensor, azimuthal velocity as the
      only streaming motion ($V_\mathrm{R}=V_\mathrm{z}=0$);
    \item 2nd component (background): observed velocity space
      ($V_\ell$,$V_b$), the centroid is fixed to the median of the
      observed velocity distribution, the velocity dispersion and the
      velocity covariance are free parameters.
\end{itemize} Table~\ref{tab:prior_disc} summarises the model parameters and their prior distributions, the number of free parameters is 6. 

We apply the fit to the subsample of 3,126 rotating RRL (see
Section~\ref{sec:kseparation} and Equation~\ref{eq:cuts}) grouped in
60 cylindrical Voronoi-cells (see Section~\ref{sec:binning}) with an
average Poisson signal-to-noise of $\approx7$.  For each region in the
$R,|z|$ plane our kinematic model provides an estimate of the
rotational velocity as well as the properties of the velocity
ellipsoid and an estimate of the background level. After our analysis,
we found a low level of contaminating background ($\approx12\%$ of
stars have $q_\mathrm{bkg}>0.7$) confirming that our subsample is a
quite clean view of the rotating disc-like RRL population.

\begin{figure}
\centering
\centerline{\includegraphics[trim = 2cm 0cm 2cm 0cm, clip=true,width=1.0\columnwidth]{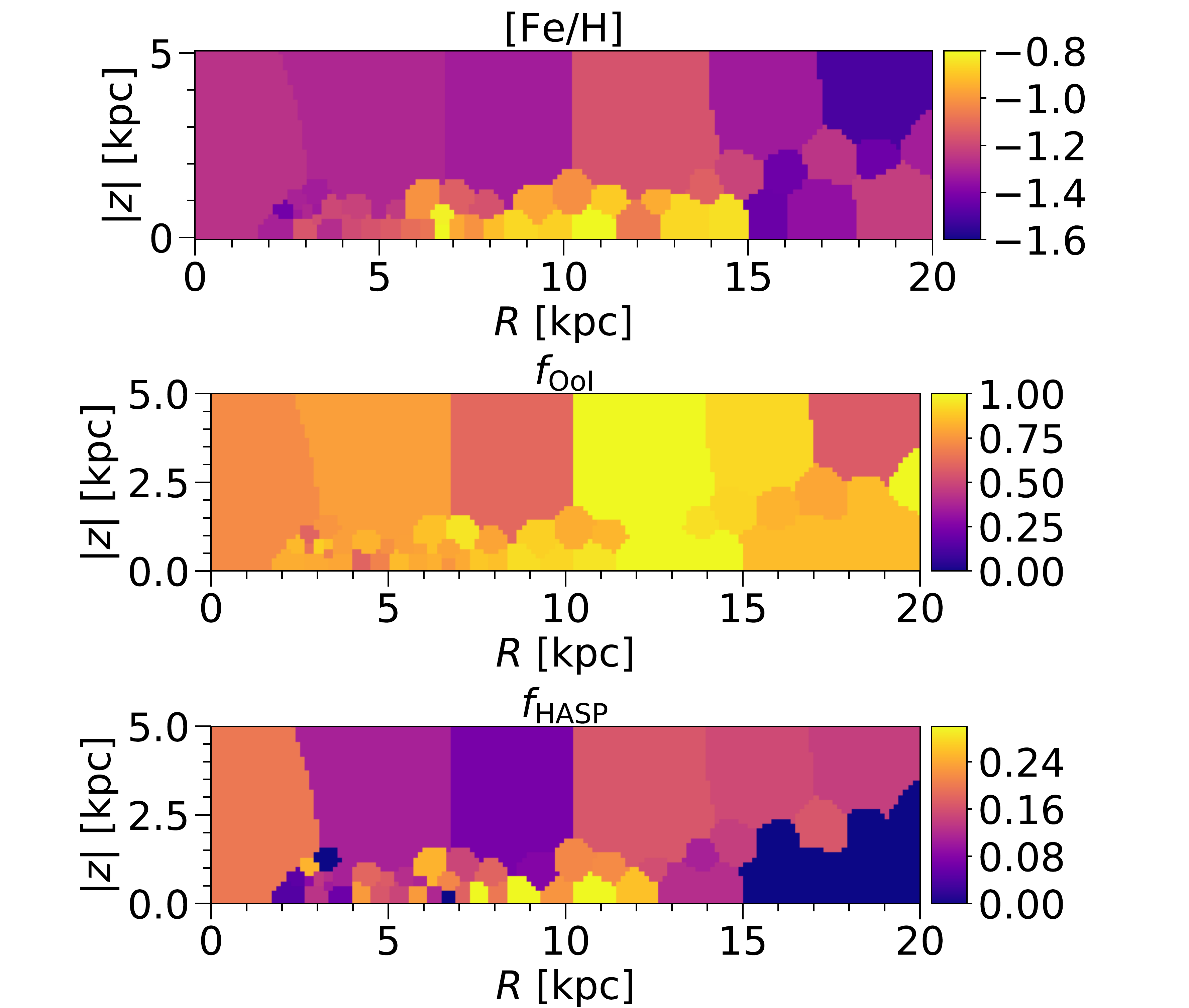}}
\caption[]{Stellar population properties of the rotating disc-like
  component in cylindrical coordinates. Top panel gives the median of
  the metallicity, middle panel shows the fraction of OoI type RR
  Lyrae, while the bottom panel presents the fraction of HASP
  stars. These maps use a subsample of the disc catalogue (see
  Section~\ref{sec:disc}) obtained considering only objects belonging
  to the SOS catalogue (1841 stars). Each bin contains at least ten
  stars. The metallicities shown in this figure have been estimated
  through Equations~\ref{eq:Met_RRab} and \ref{eq:Met_c} (see Appendix
  \ref{appendix:met}). }
\label{fig:disc_combo}
\end{figure}
\begin{figure*}
\centering
\centerline{\includegraphics[width=1.0\textwidth]{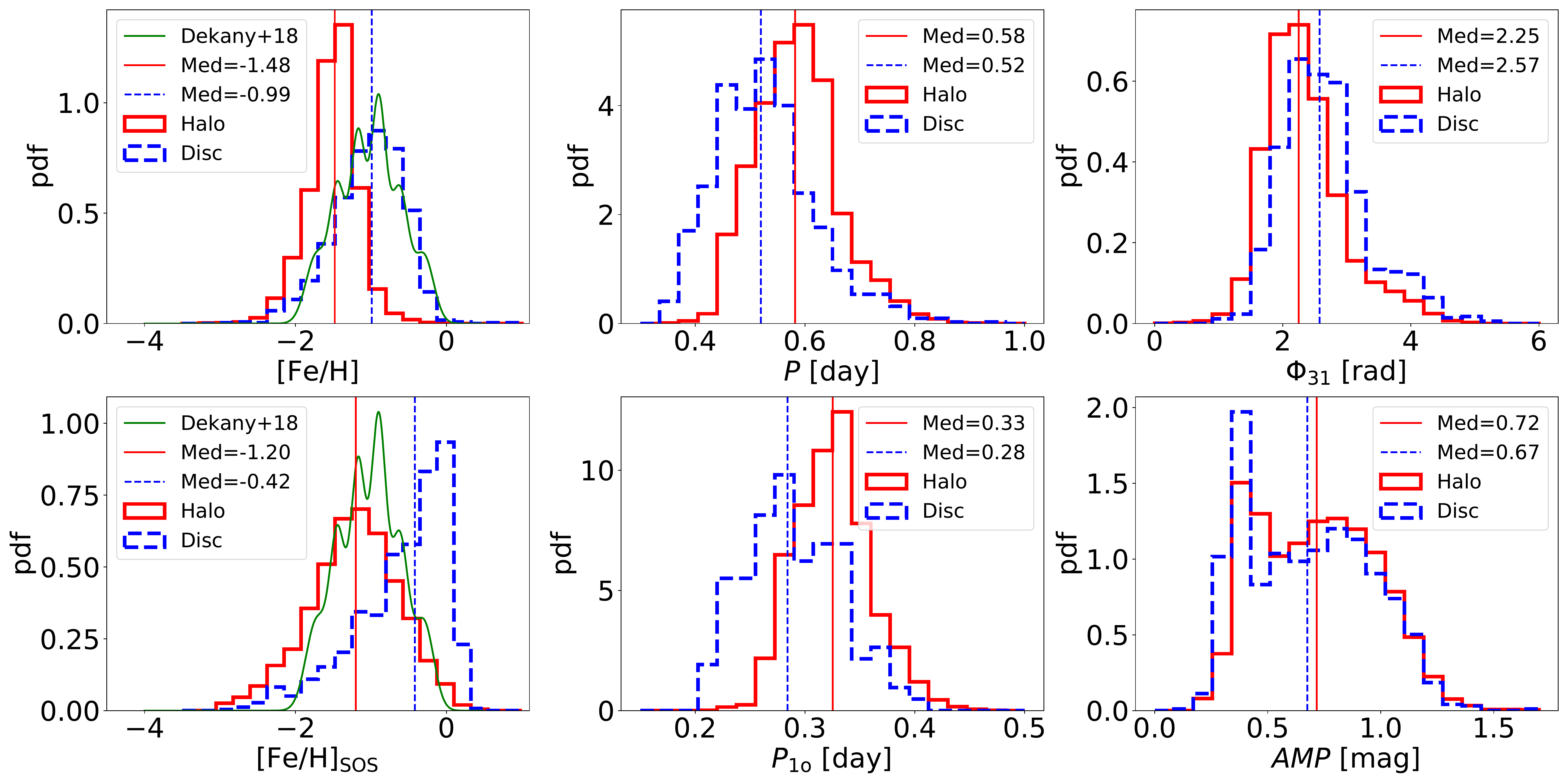}}
\caption[]{Lightcurve properties of a subsample of SOS stars in the
  Gclean catalogue (see Section~ \ref{sec:cleaning}) belonging to the
  halo (red, see Section~ \ref{sec:halo}) and the disc components
  (blue, see Section~ \ref{sec:disc}). From top-left to bottom-right
  the panels show the metallicity estimated in this work (see
  Section~\ref{sec:dist_estimate} and Appendix \ref{appendix:met}),
  the period of the RRab stars, the lightcurve phase difference
  $\Phi_\mathrm{31}$ (see Section~ \ref{sec:dist_estimate}), the
  metallicity from \gaia SOS, the period of the RRc stars and the
  lightcurve amplitude in the \gaia $G$ band. The vertical dashed
  lines give medians of the distributions.  Green curves in the
  left-hand panels show the best Gaussian Mixture Models of the
  photometric metallicity distribution of the sample of disc RRLs in
  \cite{Dekany18}.  Only stars that have estimates of both the period
  and the lightcurve phase difference have been considered for this
  plot (24,598 and 1,146 stars from the halo and disc sample,
  respectively).}
\label{fig:disc_halo_compare}
\end{figure*}

Figure~\ref{fig:disc_vrot} shows the mean azimuthal velocity (left)
and velocity dispersion (right) as a function of the Galactocentric
cylindrical radius $R$. The colour of the symbols represents their
height above the plane $|z|$. The left panel of the Figure displays a
well-behaved rotation curve traced by RRL: starting around
$V_{\phi}\approx100$ kms$^{-1}$ at distances of 2-3 kpc from the
centre of the Galaxy, it quickly rises to $V_{\phi}\approx230$
kms$^{-1}$ at $R\approx5$ kpc, and then stays relatively flat at
$5<R$(kpc)$<25$. Note that such high rotational velocities are
characteristic of the thin disc population of the Milky
Way. Overplotted on top of our measurements is the magenta line
representing the azimuthal velocity curve of the thin disc Cepheids
recently reported by \citet{Ablimit20} { and consistent with the kinematics of other thin disc tracers (e.g.\ Red Giants, \citealt{Eiler19,Lopez2014})}. In the range of Galactocentric
distances sampled by both the Cepheids and the RRL their azimuthal
velocities are in complete agreement, thus vanquishing any remaining
doubt about the nature of the fast-rotating RRL.

Stars in the Galactic disc are exposed to a variety of processes which
can change their kinematics with time. Repeated interactions with
non-axisymmetric structures such as the spiral arms, the bar and the
Giant Molecular Clouds (with additional likely minor contribution from
in-falling dark matter substructure) result in the increase of the
stellar velocity dispersion, more pronounced for older stars, often
described as Age Velocity dispersion Relation or AVR \citep[see
  e.g.][]{Stromgren1946, Spitzer1951, Barbanis1967, Wielen1977,
  Lacey1984, Sellwood1984, Carlberg1985, Carlberg1987, Velazquez1999,
  Hanninen2002,Aumer2009,Martig2014,Grand2016,Moetazedian2016,Aumer2016,Mackereth2019b,Ting2019,Frankel2020}. Most
recently, \citet{Sharma20} used a compilation of spectroscopic
datasets and \gaia DR2 astrometry to study the dependence of radial
and vertical velocity dispersions for stars with $3<R$(kpc)$<20$. They
use a combination of stellar tracers, Main Sequence Turn-Off stars and
Red Giant Branch stars whose ages are calculated using
spectro-photometric models calibrated with
asteroseismology. \citet{Sharma20} demonstrate that the stellar
velocity dispersions are controlled by four independent variables:
angular momentum, age, metallicity and vertical height. Moreover they
show that the joint dependence of the dispersion on these variables is
described by a separable functional form.

The right panel of Figure~\ref{fig:disc_vrot} compares the RRL
velocity dispersions (under the assumption of isotropy) to the median
between radial and vertical dispersion approximations obtained by
\citet{Sharma20}. Here we have fixed other model parameters to the
values most appropriate for our dataset, i.e. [Fe/H]=-1 and
$|z|=0.5$. First thing to note is that the shape of the radial
dispersion curve traced by the \gaia RRL matches remarkably well the
behaviour reported by \citet{Sharma20} for the disc dwarfs and
giants. Secondly, the RRL velocity dispersion at the Solar radius is
strikingly low, around $\approx20$ kms$^{-1}$. Overall, both the shape
and the normalisation of the RRL velocity dispersion agree well with
that predicted for a stellar population of 2 Gyr in age (orange
curve). In comparison, an older age of 10 Gyr would yield a dispersion
almost twice as large (blue curve).  Given the high azimuthal velocity
and low velocity dispersion, as demonstrated in
Figure~\ref{fig:disc_vrot} for both the Gclean and SA catalogues, we
conclude that our sample of rotating RRL is dominated by a relatively
young thin disc population.  Note that as a check, we also perform a
more detailed analysis obtaining an age estimate by fitting the
velocity dispersions with the median (radial and vertical) model
prediction from \cite{Sharma20}, considering all stars in the
disc-like subsample and their properties and errors ([Fe/H], $R$, $z$,
$V_\phi$ and $\sigma$ from the kinematic fit). This yields an age
distribution consistent with a young disc population: the peak is at
$\approx2 \Gyr$ and the wings extend from very young ages ($<1
\ \Gyr$) to 5-7 Gyr.

Our findings are in agreement with those reported in the literature
recently \citep[e.g.][]{Marsakov18, Zinn20,Prudil2020} that
demonstrate the presence in the Solar neighborhood of RRL with thin
disc kinematics and chemistry. For the first time, however, we are
able to map out the kinematics of the disc RRL across a wide range of
Galactocentric $R$ and show that their velocity dispersion behaviour
is clearly inconsistent with that of an old population. Moreover, as
demonstrated in the bottom row of Figure~\ref{fig:disc_halo}, beyond
$R\approx20 \ \kpc$ we detect prominent flare in the spatial distribution of
the disc RRL \citep[compare to e.g.][]{Lopez2014,Thomas2019}. Note
that the increase of the mean Galactic height with $R$ detected here
is gentler compared to the above studies, thus also pointing at a
younger age of these RRL in agreement with the maps presented in
\citet{Cantat2020}. Figure~\ref{fig:disc_combo} zooms in on the
rotating disc-like component and shows the properties of its stellar
population (inferred from the RRL lightcurve shapes) as a function of
cylindrical coordinates. From top to bottom, the panels show
metallicity (top), OoI fraction (middle) and HASP fraction
(bottom). Across the three panels, the disc RR Lyrae show consistent
behaviour: their metallicity, OoI and HASP fractions remain high for
$|z|<1$ kpc. For $3<R$(kpc)$<15$, radial behaviour shows no trends,
but in the very inner Galaxy, metallicity and HASP fractions
drop. Similarly, there appears to be a decrease in metallicity and
HASP fraction in the outer parts of the disc, beyond $R=15$ kpc. 
{ The
apparent central ``hole'' in the disc RRL population is consistent
with the radial offset of the metal-rich component presented in
\cite{Dekany18} and in \cite{Prudil2020}. The central depression can also be an indication of radial migration for the disc RRL population (see e.g.\ \citealt{Berardo20}). However, for our sample we can not rule out that some of the change in the inner 3 kpc at low $|z|$ is driven by the cleaning criteria applied (e.g. extinction cut) or increasing contamination from other components (bulge/bar, thick disc). 
The synchronous
change in the RRL metallicity and the HASP fraction points to the fact
that HASP objects are simply the high tail of the RR Lyrae [Fe/H]
distribution. }

Finally, let us contrast the lightcurve shapes of the halo and the
disc RRL. Figure~\ref{fig:disc_halo_compare} presents the
distributions of metallicity, period $P$, amplitude and phase
difference $\phi_{31}$ for the halo (red) and the disc (blue)
samples. We give two [Fe/H] distributions computed using two different
calibrations: the top left panel of the Figure relies on the
metallicity estimated using Equations~\ref{eq:Met_RRab} and
\ref{eq:Met_c}, while the bottom left panel employs [Fe/H] values
reported by {\it Gaia}'s SOS. Irrespective of the calibration used,
the metallicities attained by the disc RRL  are significantly
higher than those in the halo. The [Fe/H] distribution of the rotating
population exhibits a long tail towards low metallicities, but the
peak (and the median) value is higher by 0.5 (0.8) dex depending on
the calibration used. Given that the RRL  metallicities are
computed using only the period and phase difference, we expect that
both $P$ and $\phi_{31}$ distributions should show clear differences
when the halo and the disc RRL  are compared. This is indeed the
case as revealed by the middle column and the top right panel of
Figure~\ref{fig:disc_halo_compare}. The main difference is in the period
distribution: the disc RRL  have a shorter period on
average. There is also a slight prevalence of lower values of
$\phi_{31}$ while the amplitude distributions are not
distinguishable. This behavior is in happy agreement with the
properties of the disc RRL  populations gleaned from smaller local
samples \citep[see e.g.][]{Marsakov18, Zinn20,Prudil2020}.

\section{Discussion and Conclusions} \label{sec:summary}

\subsection{The unclassified stars} \label{sec:unclass}

So far, we have left out a substantial $\approx25\%$ of the total RR
Lyrae dataset as ``unclassified''. Note that according to our
definition, any sample of stars with intermediate properties, i.e. a
population that does show either a strong prograde rotation (disc) or
a zero mean azimuthal velocity (halo) would be deemed
unclassified. Here we attempt to investigate the presence of any
coherent chemo-kinematic trends amongst these leftover
stars. According to Figure~\ref{fig:maps}, the bulk of this
unclassified population gravitates to the centre of the Milky Way and
sits close to the plane of the disc.

\begin{figure*}
    \centering
\centerline{\includegraphics[trim= 0cm 6cm 0cm 6cm,width=1.0\textwidth]{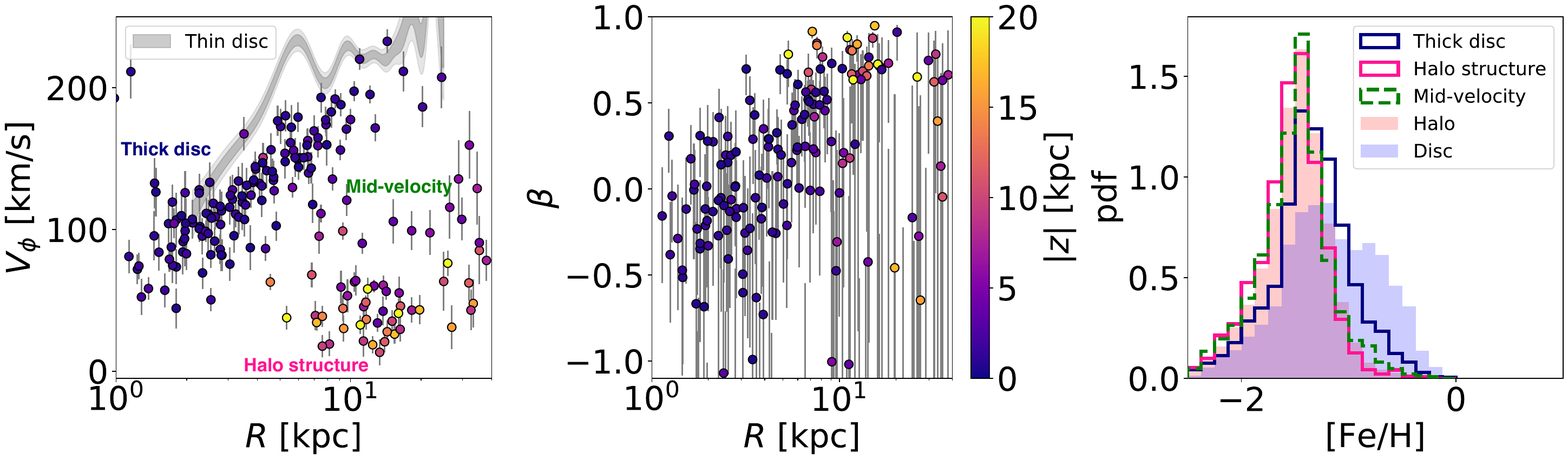}}
    \caption{Chemo-kinematic analysis of the unclassified subsample
      (see Section~ \ref{sec:kseparation} and
      Figure~\ref{fig:maps}). {\it Left}: rotational velocity as
      function of the cylindrical radius, the grey bands show the
      GP-interpolation of the rotational velocities obtained for the
      rotating disc-like component (see
      Figure~\ref{fig:disc_vrot}). {\it Centre}: anisotropy parameter
      as a function of the cylindrical radius. The color map in the
      left and middle panels indicates the median value of the
      absolute value of $z$, the points and the error bars indicate
      the median values, the 16th and 84th percentile correspondingly
      of the a-posteriori distribution obtained for each bin. {\it
        Right}: metallicity distribution for the SOS stars in the
      unclassified subsample, the unfilled blue histogram contains the
      unclassified stars with thick-disc like kinematics, the unfilled
      magenta histogram shows the distribution for unclassified stars
      with halo like kinematics while the unfilled dashed-green
      histogram contains unclassified stars in bins with intermediate
      azimuthal velocity ($\approx 100 \ \kms$). As comparison, the
      blue and red filled histograms show the metallicity distribution
      of the stars belonging to the the halo-like and disc-like
      components (see Section~ \ref{sec:kseparation} and
      Figure~\ref{fig:disc_halo_compare}). The metallicities shown in
      this figure have been estimated through
      Equations~\ref{eq:Met_RRab} and \ref{eq:Met_c} (see Appendix
      \ref{appendix:met}).}
    \label{fig:uns}
\end{figure*}

Figure~\ref{fig:uns} presents the results of the kinematic
modelling\footnote{The fit parameters and their prior distributions
  are the same of the anisotropic halo component summarised in Table~
  \ref{tab:halo} but with $L_\mathrm{r}\sim\delta(0)$. The total
  number of free parameter is 3.}  of the hitherto unclassified RRL
stars. The left panel of the Figure shows the mean azimuthal velocity
as a function of Galactocentric $R$ with the colour-coding
corresponding to $|z|$. Two main groups are immediately
apparent. First, between 1 and 10 kpc from the Milky Way's centre, at
low heights, there exists a population of RRL rotating with speeds
lagging behind the thin disc by some $\approx50$ kms$^{-1}$ which we
attribute to the thick disc population. It is interesting to note that
a hint of the presence of a population with thick-disc like kinematics
is already shown in Figure~\ref{fig:disc_vrot}: approximately at the
Sun position we can identify a clear vertical gradient of the
azimuthal velocity. In particular, the $V_\phi$ of the point with
$|z|\approx2 \kpc$ is consistent with the thick-disc velocities shown
in Figure~\ref{fig:uns}.

Additionally, beyond $R>10$ kpc and $|z|>10$ kpc above the plane,
another barely rotating population is discernible - most likely
belonging to the halo. There is also a small number of bins that
display kinematical properties in between the thick disc and the
halo. Interestingly, the halo portion of the unclassified RRL exhibit
high orbital anisotropy $\beta\approx0.8$ as evidenced in the middle
panel of Figure~\ref{fig:uns}. This would imply that much of this halo
substructure is attributable to the \gaia Sausage. This is in
agreement with the earlier claims of \citet{Simion2019} who connect
the Virgo Overdensity and the Hercules Aquila Cloud to the same merger
event. In fact, in Figure~\ref{fig:maps}, traces of both the VOD and
the HAC are visible amongst the unclassified RRL stars. Note that
assigning the slowly-rotating portions of the halo to the GS debris
cloud would increase the net angular momentum of this radially-biased
halo component. The bins dominated by the thick disc stars have
$\beta\approx0$ with a mild increase with radius $R$. It is curious to
see that the slowly rotating RRL population is limited to $R<12$ kpc
as has been seen in many previous studies
\citep[e.g.][]{Bovy2012,Hayden2015,Bland-Hawthorn2019,Grady2020}
supporting the picture where rather than just thick, this is an inner,
old disc of the Galaxy.

The right panel of Figure~\ref{fig:uns} presents the metallicity
distributions of the halo (unfilled magenta), thick disc (unfilled
blue) and intermediate $v_{\phi}$ (green dashed) populations amongst
the previously unclassified RRL. These can be compared to the
halo (filled light red) and thin disc (filled light blue) [Fe/H]
distributions. Reassuringly, the bits of halo substructure with slight
prograde motion have the [Fe/H] distribution indistinguishable from
the that of the halo's sample. The thick disc displays metallicities
that are on average lower than the thin disc's but not as low as in
the halo. Based on the chemo-kinematic trends amongst the
`unclassified' stars, we conclude that the majority $\approx70\%$
belong to the Milky Way's thick disc, while the remaining
$\approx30\%$ are part of the halo substructure, which displays the
prevalence for prograde motion and high orbital anisotropy.

\subsection{Tests and caveats} \label{sec:tests}

The results of this work rely on a number of assumptions. In this
section we quantify the impact of some of the possible systematics,
repeating the analysis of the halo and the disc kinematics (see
Section~ \ref{sec:kfit}, Section~\ref{sec:halo} and
Section~\ref{sec:disc}).

One of the principal ingredients of our modelling is the distance
estimate for the RRL stars in our sample.  We investigate the role of
a potential distance bias using the SOS metallicity estimate instead
of the one presented in this paper in Equation~ \ref{eq:Met_RRab} and
Equation~ \ref{eq:Met_c} (see Appendix \ref{appendix:met}). Moreover,
we test the effect of assuming a constant absolute magnitude,
$M_\mathrm{G}=0.64\pm0.24$ (see Appendix \ref{appendix:met}), in
Equation~ \ref{eq:amag}. 
{
 We  are happy to report that all main conclusions of our analysis remain unchanged.  The radial profile of the fitted halo and disc properties are all within 1$\sigma$ of our fiducial results and we do not find any significant systematic
differences between the outcomes.
}

{ The separation of the halo and disc component relies on a selection cut based mainly on the a-posteriori likelihood to belong to the non-rotating halo component (see Section`\ref{sec:kseparation}). We do not repeat the kinematic analysis  for different $q_\mathrm{halo}$-thresholds, but looking at Figure~\ref{fig:uns} the result of such an experiment is easily extrapolated. Increasing the value of the disc $q_\mathrm{halo}$-cut we include more and more of thick disc stars (that are larger in number) lowering the rotational velocity, increasing the velocity dispersion and lowering the metallicity. This does not change our conclusions but just hides the subdominant  thin-disc-like component under a large number of stars belonging to a different kinematic component.}

Part of the halo analysis relies on splitting the stars into spherical bins,
however, the inner stellar halo is known to be flattened (see
e.g.\ \citealt{Deason11,Xue15,Das16,Iorio18,Iorio19}). We repeat the
kinematic fit of the halo subsample using elliptical bins instead,
tuned on the ellipsoidal shape described in \cite{Iorio19}. Comparing
the outcomes of the spherical and elliptical analysis we do not find
any significant differences. { Moreover,  we  perform an alternative  
analysis binning the volume in cylindrical coordinates, so that the results are independent on the assumption of spherical or elliptical symmetry (but still dependent on the azimuthal symmetry, see below). The results of the cylindrical analysis are qualitatively in agreement with the 1D radial profile obtained assuming spherical symmetry (see Figure~ \ref{fig:halo_comb} and Figure~ \ref{fig:halo_comb_vphi}).}

We test the assumption of the four-fold symmetry repeating our
analysis considering only stars located in a given Galactic quadrant,
i.e. we select stars based on their Galactic azimuthal angle.
{ We do not detect any significant difference or systematic offset in the fitted halo and disc parameters (within 1$\sigma$ of our fiducial results)}, except for the azimuthal
velocity of the radial component of the halo (see Section~
\ref{sec:halo_kin}). This parameter shows a significant offset
depending on the considered quadrants: in the Galactic semi-plane not
containing the Sun ($90^\circ<\Phi<270^\circ$), the average azimuthal
velocity is negative ($V_{\phi,\mathrm{rad}}\approx-25 \ \kms$), while
in the other portion of the Galaxy $V_{\phi,\mathrm{rad}}$ is just
slightly higher than 0, except in the innermost part where it rises up
to $30-40 \ \kms$. The final velocity profile showed in
Figure\ \ref{fig:halo_comb_vphi} is approximately the weighted mean
(there are more stars in the quadrants closer to the Sun) of the
$V_{\phi,\mathrm{rad}}$ profiles obtained considering the four
different quadrants. Although we cannot exclude the presence of real
asymmetries or hidden halo subcomponents, it is more likely that this
difference is driven by the distance biases present (see
e.g.\ \citealt{AllegedDuality,Schonrich12}).  Indeed, the velocity
offset is dependent on the distance from the Sun with more distant
quadrants showing a larger deviation from $V_{\phi,\mathrm{rad}}=0$.
Curiously, the velocity offset is not present in the isotropic
component, however $V_{\phi,\mathrm{iso}}$ is in general less
constrained. In that case, the random errors are likely dominating the
error budget reducing the effect of the systematic offset.
 
The results for the thin disc are obtained assuming isotropy, hence we
repeat the fit leaving the three components of the velocity ellipsoid
free ($\sigma_\mathrm{R},\sigma_\mathrm{z},\sigma_\mathrm{\phi}$). We
also model the non-diagonal terms of the correlation matrix as
nuisance parameters. The results are consistent with those shown in
Figure\ \ref{fig:disc_vrot}, in particular the three velocity
dispersions agree within the errors confirming that our assumption of
isotropy is supported by the data.  However, we do expect a certain
degree on anisotropy in the disc
($\sigma_\mathrm{R}>\sigma_\mathrm{z}$, see
e.g.\ \citealt{Sharma20,Katz18}). The reason why we do not detect the
velocity dispersion anisotropy in our data is unclear. It is possible
that we are introducing some selection bias in the kinematic
decomposition (Section~ \ref{sec:kseparation}) as we force the
rotating component to be isotropic. It could also be that the
differences are washed out by the noise in our data and by the
limitation of our analysis. In particular, most of the stars in the
rotating subsample have small $z$ (see Figure\ \ref{fig:maps}), hence
$V_b$ is almost directly mapping $V_\mathrm{z}$ while the other two
velocity components are harder to constrain.  Despite this possible
issue about the velocity dispersion, the model parameters of the
rotating component (azimuthal velocity and velocity dispersion, see
Section~ \ref{sec:disc}) are relatively insensitive to any of the
tested variations, therefore the association of this component with
the kinematic thin disc is robust.

Concerning the chemical analysis it is important to stress that it is
based on photometric metallicities (see Appendix
\ref{appendix:met}). As already noted by \cite{Clementini} and
\cite{Cacciari05}, such photometric estimates are not suited to
describe individual metallicities but rather the average metal
abundance of a population. Moreover, as shown in
Figure\ \ref{fig:disc_halo_compare}, the photometric metallicity can
differ significantly between different calibrations. Most of our
analysis is based on the comparison between metallicity distributions
of groups of stars (see Figure\ \ref{fig:halo_comb_met} and
Figure\ \ref{fig:disc_halo_compare}), hence the results should be robust
despite the limitation imposed by the use of photometric
metallicities. Concerning the rotating disc-like component it is
evident that the metallicity is on average higher with respect to the
halo.  However, given the uncertainty of the photometric metallicities
it is hard to constrain the real average metallicity of this
population. As discussed in Appendix \ref{appendix:met}, we notice
that our photometric estimate seems to underestimate high
metallicities, on the contrary the metal abundance reported in the SOS
catalogue tends to overpopulate the high metallicity end of the [Fe/H]
distribution.  Therefore, we conjecture that the true average value is
somewhere between our estimate ([Fe/H]$\approx-1$) and the higher
value estimated in the SOS catalogue ([Fe/H]$\approx-0.4$). 
{ Interesting, we notice that the high resolution spectroscopic datasample of field RRL from \cite{Magurno18} shows a clear metal-rich component, ranging between [Fe/H]$\approx-0.5$ and [Fe/H]$\approx0.2$,  in the metallicity distribution (see Figure~12 in \citealt{Fabrizio2019} and Figure~\ref{fig:compare_savino} in Appendix \ref{appendix:met}.) }

Recently, \cite{Berardo20} pointed out that our comparison with the \cite{Sharma20} models could be biased toward younger age because our sample is kinematically selected. However, we stress that the \cite{Sharma20} models take into account the kinematics through the vertical angular momentum parameter, $L_\mathrm{z}$. Indeed,  at a given age, they predict smaller velocity dispersions for larger $L_\mathrm{z}$,  this is an expectation of the model not an effect of a selection bias. It is important to note that in our case we can associate  $L_\mathrm{z}=V_\phi R$ to  each star in a bin (see Fig.\ \ref{fig:disc_combo}), so the selection on $V_\phi$ (selecting small $q_\mathrm{halo}$) as well on $z$ (see Eq.\ \ref{eq:cuts}) are not introducing any bias since they are both parameters of the \cite{Sharma20} models and the only free parameters of our analysis is the population age. 

\cite{Berardo20} conclude that the presence of a population of old RRL in the thin disc can be easily accommodate considering an early co-formation of thin and thick discs. This can surely be the case, but we stress once again that the progenitors of metal-rich RRL ([Fe/H]$>-1$) need a significant mass loss to reach the instability strip regardless of their age.

\subsection{The bulge/bar}

The closest the stars in our sample get to the Galactic centre is
$\approx1.3$ kpc. Combined with the restriction on the dust reddening
which eliminates low latitudes, this implies that the Milky Way's bar
and bulge are mostly excluded from our study. As of today, OGLE
\citep[e.g.][]{Soszynski2014} and VVV \citep[][]{Dekany2013} surveys
provide much better view of the RR Lyrae properties in the heart of
our Galaxy. The structure and the metallicity distribution of the
bulge region as traced by RR Lyrae appear complex and puzzling and
agreement is yet to be reached as to the exact interplay of distinct
Galactic components here
\citep[][]{Pietrukowicz2015,Kunder2016,Dekany18,Prudil2019a,Prudil2019b,Kunder2020,Du20}. The
bulge tangled mess might well have reached into our sample for stars
with distances $R<4$ kpc from the Galactic centre, but their numbers
are low and their (potential) contribution does not change any of the
conclusions reported here.

\subsection{Conclusions}

We use \gaia DR2 proper motions to identify individual Galactic
components amongst RRL, pulsating horizontal branch stars, usually
assumed to be mostly old and metal-poor. Following the ideas recently
highlighted in \citet{Wegg19}, we assume four-fold symmetry to extract
the properties of the 3D velocity ellipsoid as a function of
Galactocentric distance $R$ and height $|z|$. The \gaia DR2 RRL
catalogue is dominated by stars with halo kinematics ($\approx70\%$),
i.e. those with little prograde rotation. Some $\approx5\%$ of the RR
Lyrae have fast azimuthal velocities, $v_{\phi}\approx220-230$
kms$^{-1}$, while the remaining $\approx25\%$ are unclassified,
i.e. have kinematic properties intermediate between the halo and the
thin disc. We further demonstrate that the halo sample contains at
least three distinct sub-populations. The unclassified sample is
dominated by the thick disc stars with a small addition of a mildly
prograde halo debris.

Between 50\% and 80\% of the halo RRL stars with $5<R$(kpc)$<25$
belong to the radially biased ($\beta\approx0.9$) non-rotating (or
perhaps slowly rotating) structure known as the \gaia Sausage, left
behind by an ancient merger with a massive dwarf galaxy \citep[see
  e.g.][]{Deason2013,Belokurov2018,Haywood2018,Deason2018,Helmi2018,Mackereth2019,Lancaster2019,Fattahi2019}. The
remainder of the halo is much more isotropic and probably contains a
mixture of stars accreted from lower-mass satellites.  The \gaia
Sausage component exhibits little angular momentum and a strong
bimodality in the radial velocity
\citep[see][]{Lancaster2019,Necib2019}. We model the radial velocity
distribution of the \gaia Sausage with two Gaussians separated by
$2L_r$ and show that the amplitude of the radial velocity separation
is a strong function of the Galactocentric distance $R$. $L_r$ peaks
around $3<R$(kpc)$<5$, the distance, we conjecture, which marks the
location of the pericentre of the GS, while its apocentre is close to
$R\approx25$ kpc where $L_r$ drops to 0 kms$^{-1}$. The GS debris is
distinct from the rest of the halo not only kinematically but also in
terms of the lightcurve shapes of the constituent RRL. Compared to the
isotropic halo, the GS RRL boast a higher fraction of Oosterhoff Type
1 objects. Beyond $R\approx10$ kpc, the GS stars are more metal-rich
than the isotropic halo, and additionally exhibit a higher fraction of
the HASP RRL \citep[in agreement with e.g.][]{BelokurovHASP}
supporting the massive merger scenario. However, within 10 kpc, there
exists a subset of the isotropic halo RRL whose metallicity and HASP
fraction is even higher than those in the GS. We conjecture that these
inner metal-rich and HASP-rich RRL were born in situ
\citep[representing the population previously seen in
  e.g.][]{Nissen2010,Bonaca2017,Haywood2018,
  DiMatteo2019,Gallart2019,Belokurov2020}.

We are not the first to detect RRL stars with disc kinematics
\citep[see][]{Kukarkin1949,Preston1959,Taam1976,Layden94,Layden1995a,Layden1995b,MateuDisc,
  Marsakov18,Marsakov2019,Prudil2020,Zinn20}. Note however, that these
previous studies have been mostly limited to the Solar
neighborhood. Here for the first time we map out the kinematics of the
disc RRL over the entire extent of the disc, i.e.
$3<R$(kpc)$<30$. The RRL with the fastest azimuthal speeds in our
sample follow closely the thin disc behaviour, both in terms of their
rotation curve and the evolution of the velocity dispersion. Using the
recent models of the velocity dispersion obtained for conventional
thin disc tracers such as MS and RGB stars by \citet{Sharma20} we
place strong constraints on the typical age of the thin disc RR
Lyrae. The thin disc traced by the \gaia RRL is very cold and can not
be more than $\approx 5$ Gyr old. Moreover, we demonstrate that the
thin disc RRL ought to be significantly more metal-rich compared to
their halo counterparts, in agreement with the earlier studies
mentioned above. The thick disc RRL are also detected as part of our
study. These stars do not rotate as fast and hence are placed in the
``unclassified'' category. Careful examination of these stars with
intermediate kinematic properties reveal that in bulk, they are
denizens of the thick disc. Their lightcurve shapes indicate that they
only slightly more metal-rich compared to the halo. Curiously, the
kinematically-selected thick disc RRL do not tend to reach beyond
10-12 kpc from the Galactic centre, in agreement with the theories of
the thick disc formation.

We draw attention to the fact that the existence of young and
metal-rich RRL stars in the thin disc can not be easily reconciled
with the predictions of the accepted, single-star evolutionary model:
metal-rich young progenitors require un-physically high mass
loss. Perhaps, instead we have discovered an army of RR Lyrae
impostors (akin to BEPs) produced via mass transfer in binary systems.

\section*{Acknowledgements}
The authors thank the anonymous referee for suggestions that helped to improve the manuscript.
We are grateful to M\'arcio Catelan, Gisella Clementini, Alessandro Savino  and Leandro Beraldo e Silva for the thoughtful comments they supplied on the earlier version of the manuscript. We thank Jason Sanders, GyuChul Meyong,
Eugene Vasiliev, Wyn Evans and the other members of the Cambridge
Streams group for the stimulating discussions at the early stage of
this work.  We thank Iulia Simion for useful discussions and the for
her help in the cross-match of the Liu+20 dataset with Gaia RR
Lyrae. We thank Yang Huang for sharing the Liu+20 dataset. GI wish
to thank Nicola Giacobbo for inspiring discussions.  During his period
in UK, GI was supported by the Royal Society Newton International
Fellowship. VB is grateful to Nat\`alia Mora-Sitj\`a for the careful
proof-reading of the manuscript. This work has made use of data from
the European Space Agency (ESA) mission Gaia
(\url{https://www.cosmos.esa.int/gaia}), processed by the Gaia Data
Processing and Analysis Consortium (DPAC,
\url{https://www. cosmos.esa.int/web/gaia/dpac/consortium}). Funding
for the DPAC has been provided by national institutions, in particular
the institutions participating in the Gaia Multilateral Agreement.
The research has made use of the NASA/IPAC Extragalactic Database
(NED) which is operated by the Jet Propulsion Laboratory, California
Institute of Technology, under contract with the National Aeronautics
and Space Administration.

\section*{Data Availability}
The data underlying this article  are available in Zenodo at \url{http://doi.org/10.5281/zenodo.3972287}.


\bibliographystyle{mnras}
\bibliography{main} 

\appendix
\section{Photometric metallicity estimate} \label{appendix:met}

Most of the stars in the SOS \gaia catalogue have photometric
metallicities \citep{Clementini} estimated through the non-linear
relation by \cite{Nemec13}. The \cite{Nemec13} relation has been
fitted to a small sample of stars and it does not seem to generalise
well enough on larger sample. In particular, it assigns to a group of
RRL with intermediate-large periods and large $\Phi_{31}$ high
metallicities ($[Fe/H]\gtrsim-0.5$) that are likely artefacts (see
e.g.\ Fig.\ \ref{fig:compare_savino}). Moreover, the relation is based
on the Kepler magnitude band and a number of auxiliary relations have
to be used to translate the $\Phi_{31}$ from the original band to the
\gaia one \citep{Clementini}; additionally the value of $\Phi_{31}$
can change if we use a different number of harmonics to decompose the
light curve.  For all these reasons, we decide to find a relation
based solely on the light curve properties reported in the \gaia SOS
catalogue.  For the purpose of our analysis we cross-matched the
subsample of RRab stars with complete SOS light curve information in
our Gclean catalogue (see Sec.\ \ref{sec:sample}) with different
spectroscopic sample of RRab stars with spectroscopic metallicity
estimate: \cite{Layden94} (84 stars), \cite{Marsakov18} (76 stars),
\cite{Nemec13} (21 stars), \cite{Zinn20} (149 stars, mostly based on
the sample by \citealt{Dambis13} containing also the 84 stars in
\citealt{Layden94}).  Concerning the RRc stars we follow
\cite{Nemec13} considering the RRL in globular clusters (50 stars)
assigning them the metallicity of the cluster they belong. We use the
catalog of \gaia objected associated with Globular Clusters in
\cite{HelmiGaia}, while the Globular Cluster metallicity are taken
from \cite{Globularcluster}. We consider the old
\cite{Globularcluster} compilation because the metallicities are
reported in the \cite{Zinn84} metallicity scale instead of the
\cite{Carretta09} scale used in the more recent \cite{Harris10}
catalogue. The \cite{Zinn84} scale is the same metallicity scale of
spectroscopic catalogs and the absolute magnitude-metallicity relation
used in this work has been calibrated on this same scale
\citep{Muraveva18}.
 
\begin{figure}
\centering
\centerline{\includegraphics[width=1.0\columnwidth]{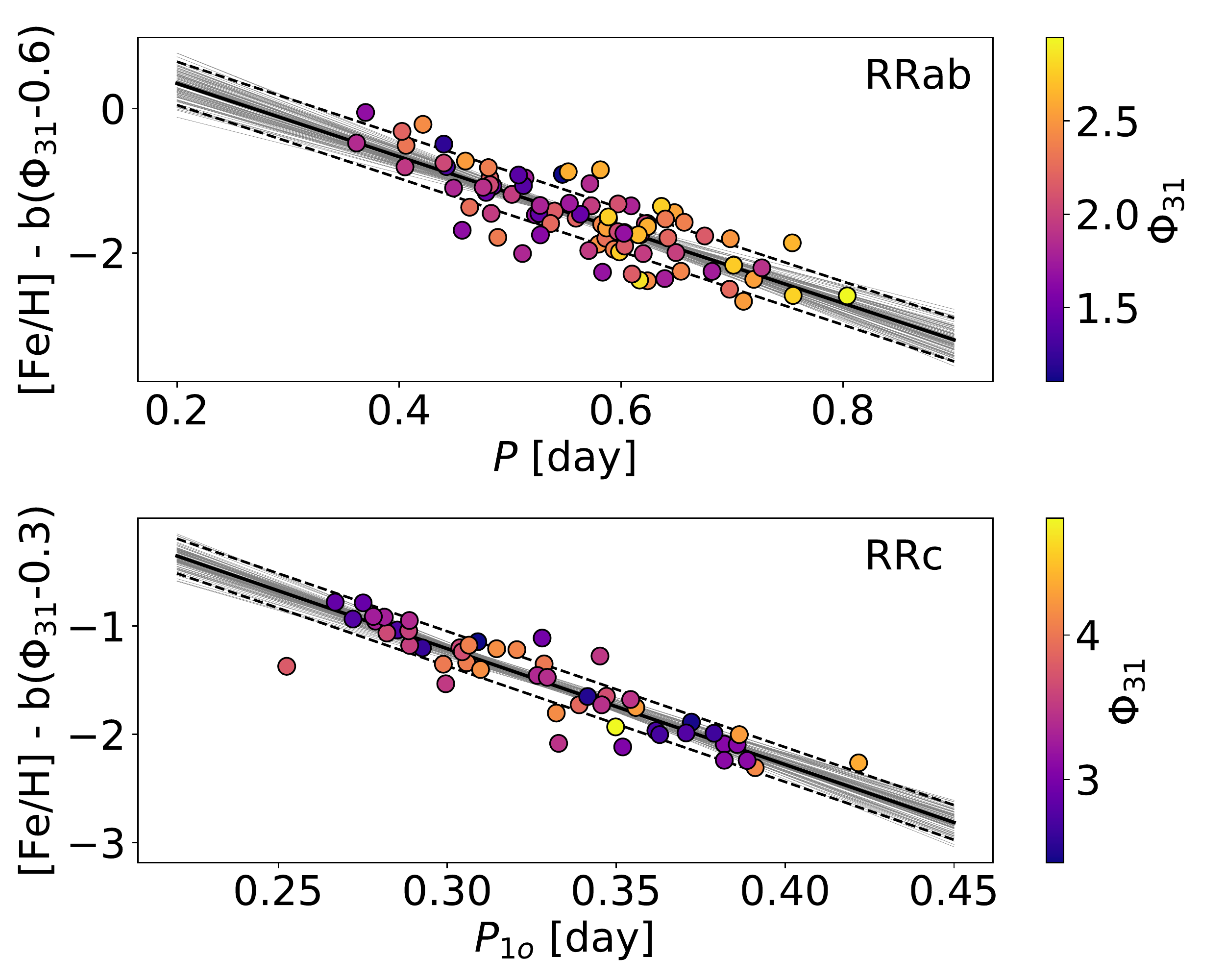}}
\caption[]{Best fit linear relation $[Fe/H]\propto a\times~P + b\times
  \Phi_{31}$ for RRab (top panel) and RRc stars (bottom panel). The
  spectroscopic metallicities are from \cite{Layden94} and
  \cite{Globularcluster} for RRab and RRc stars, respectively. Periods
  and phase difference $\Phi_{31}$ values are from the SOS \gaia
  catalogue. The solid black lines show the median of the posterior
  distributions of the relations, while the gray lines are randomly
  sampled from the same distributions. The black dashed lines indicate
  the intrinsic scatter. The best fit relations are given in
  Equations~\ref{eq:Met_RRab} and \ref{eq:Met_c}.}
\label{fig:best_fit_met}
\end{figure}

We perform a large number of tests using both linear (e.g.\ \citealt{Jurcsik96,Smolec05}) and non linear relations (e.g.\ \citealt{Nemec13}) and investigating different combinations of light curve and stellar properties.
Initially, we evaluate the feature relevance through a random forest regression of the metallicity using the \texttt{scikit-learn} python module \citep{scikit-learn}. In practice we consider as feature: the period $P$ (fundamental period  for RRab and first overtone period for RRc), the phase difference between the third  or second light curve harmonics with respect to the fundamental one, the amplitude, the ratio between the amplitude of third or second light curve harmonics with respect to the fundamental one and the stellar color. In order to check possible biases and artefacts we also add the number of \gaia observations, the mean $G$ magnitude and the $RUWE$ to the group of features.
For both RRab and RRc samples the most relevant feature is by far the period $P$, followed by the phase difference $\Phi_{31}$. We do not use the random forest method to estimate the metallicity since our training sample is relatively small and, considering the large number of parameters involved, it is very likely to produce a significant variance or overfit problem. Instead we fit the relations  using a Bayesian approach taking into account the uncertainties of all the used features. In each tested relation we consider also the presence of an intrinsic scatter. We sample the posterior of the relation parameters exploiting the Hamiltonian MCMC technique making use of the python module \texttt{PYMC3} \citep{pymc3}. The performance of the various relations are analysed considering: $i$) fit residuals, $ii$) comparison with metallicities of RRL stars in Globular clusters (association with GC from \citealt{HelmiGaia},  metallicites estimate from \citealt{Globularcluster}), $iii$) comparison with the spectroscopic metallicities of the RRL stars in the solar neighbours,the halo and the bulge   taken from the crossmatch with the \cite{Magurno18}, \cite{Liu20} and  \cite{Savino20} samples (see Fig.~\ref{fig:compare_savino}), $iv)$ comparison of the distance moduli derived using the $M_G-[Fe/H]$ relation by \cite{Muraveva18} with the distance moduli of  the Magellanic Clouds\footnote{We used the median of the distance moduli estimates taken from NED (NASA/IPAC Extragalactic Database, \url{http://ned.ipac.caltech.edu}).}. We conclude that the optimal fit, both for RRab and RRc stars,  is obtained with a linear relation with $P$ and $\Phi_{31}$, very little improvements can be obtained using non-linearity or adding parameters to the relation. As already noted by \cite{Jurcsik96,Smolec05,Nemec11}, the major issue is a moderate systematic trend of the residuals as a function of the spectroscopic metallicities: the relation tends to overestimate (underestimate) the metallicity at the metal-poor (metal-rich) end. Anyhow, this problem is  present with the same significance also with more complex models. This is likely due to the lack of calibrators at both ends of the metallicity distribution.
Among the various samples of RRab, the results of the fit are very similar except for the Nemec sample, but it  contains a small number of stars covering a narrower range of metallicites with respect to the other samples. Therefore, we adopt as final relations (Equation \ref{eq:Met_RRab} and  \ref{eq:Met_c}), the linear relation in $P$ and $\Phi_{31}$ obtained with the \cite{Layden94} sample (for RRab stars). This choice  is motivated by the fact that it is not a collection of different catalogues and it reports a metallicities uncertainty for each star. Fig.\ \ref{fig:best_fit_met} shows the best-fit relations. 
{The metallicity interval  of the fit training set ranges from -2.51 to 0.08 for the
  RRabs stars and from -2.37 to -0.55 for the RRc stars. Only a very small portion (mostly RRc stars) of our Gclean sample (see Sec.\ \ref{sec:cleaning}) has  metallicities extrapolated outside these ranges: 396 at the metal-poor tail (93 RRab, 303 RRc, 295 in the halo subsample, 6 in the disc subsample), 105  at the metal-rich end (26 RRab, 79 RRc, 15 in the halo subsample, 42 in the disc subsample). These numbers are small enough to have negligible effects on our outcomes as confirmed by the results obtained with the SA sample (see e.g. Fig. \ref{fig:disc_vrot} and Fig.\ \ref{fig:halo_comb}) that contains only $0.3\%$ of stars with extrapolated metallicities. Moreover, the fit procedure ``naturally" assigns larger errors to extrapolated metallicities and the implemented linear function limits uncontrolled behaviour outside the range of calibrators.}

Compared to the photometric metallicities reported in the \gaia SOS catalogue our estimate perform better both on estimating the absolute magnitude of the stars in the Magellanic Clouds (using the $M_G-[Fe/H]$ relation by \citealt{Muraveva18}) and compared to the RRL sample of spectroscopic metallicity obtained by  \cite{Savino20}, \cite{Liu20} and \cite{Magurno18}. Fig.\ \ref{fig:compare_savino}  shows that the distribution of  SOS photometric metallicities significantly differs from the spectroscopic ones in both shape and centroid position (see also \citealt{Hajdu19}). In particular, considering the bulge sample, the SOS distribution peaks at a very metal-rich value of $[Fe/H]\approx-0.5$,  while the peak of the spectroscopic metallicity is $[Fe/H]\approx-1.5$. The photometric metallicity estimated with our relation shows a more similar distribution with a coincident but narrower peak. The narrow distribution of the photometric metallicities is due to the already discussed problem of overestimating/underestimating the metallicities at the edge of the distribution. Considering the \cite{Liu20} sample, our metallicity distribution is slightly offset from the spectroscopic distribution, but overall the distribution widths are very similar. On the contrary, the SOS distribution is much more spread containing a significant number of metal-rich stars ($[Fe/H]>-1$).
{The  peak of  the distribution of our photometric metallicities is consistent  with the peak of the high resolution spectroscopic metallicities  in \cite{Magurno18}, but in this case the differences in the tails are more significant. For the same sample, the SOS photometric metallicities cover the same range of the spectroscopic metallicities, but their distribution is much flatter without a clear peak and  with an over-abundance of very metal-rich stars.}

\begin{figure}
\centering
\centerline{\includegraphics[width=0.8\columnwidth]{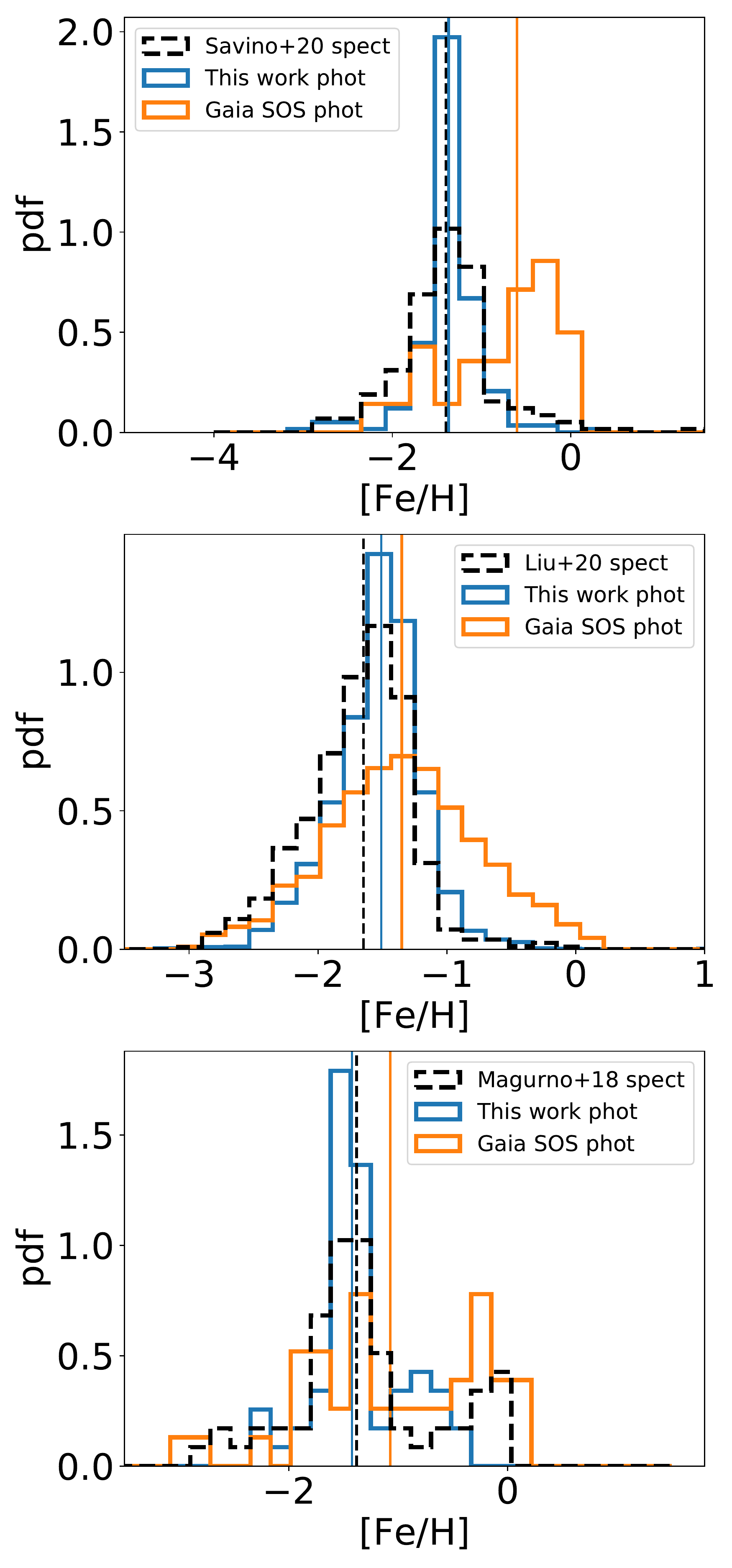}}
\caption[]{Comparison between the distribution of photometric (this
  work, blue; \gaia SOS orange) and spectroscopic (dashed-black)
  metallicity values for two samples of RRL. Top panel: cross-match
  between the bulge RRL sample in \cite{Savino20} and \gaia SOS with
  lightcurve information (212 stars). Middle panel: cross-match
  between the RRL sample (mostly in the halo) from \cite{Liu20} and
  \gaia SOS with lightcurve information (3153 stars). { Bottom panel: cross-match between the RRL sample (local field) from \cite{Magurno18} and the \gaia SOS with lightcurve information (64 stars).}
  Vertical lines
  indicate the median of each distribution.}
\label{fig:compare_savino}
\end{figure}

Finally, we test that the use of the constant absolute magnitude $M_G=0.64\pm0.25$ for both RRab and RRc stars (see e.g.\ \citealt{Iorio19}) is a good approximation when light curve properties are not available.The associated error $\delta M_G=0.25$ is a robust  and conservative  estimate  that can absorb both random  and systematic uncertainties (e.g.\ RRL type, metallicity) giving a error on heliocentrinc distance of about $13\%$.

\section{Rotation Matrix} \label{appendix:rotation_matrix}

The  rotation matrix $\mathbf{R}$ to pass from velocities in Spherical $\bm{V}_\mathrm{sph}=(V_\mathrm{r},V_\mathrm{\theta},V_\mathrm{\phi})$ or Cylindrical $\bm{V}_\mathrm{cyl}=(V_\mathrm{R},V_\mathrm{z},V_\mathrm{\phi})$ Galactocentric coordinates to the velocities in the observed frame of reference $\bm{V}_\mathrm{sky}=(V_\mathrm{los},V_\ell,V_b)$ can be obtained  with the matrix product 

\begin{equation}
\mathbf{R}= \mathbf{R}_\mathrm{c}     \cdot \mathbf{R}_\mathrm{s,sph/cyl}    
\end{equation}
where $\mathbf{R}_\mathrm{c}$ is the rotation matrix  to pass from the Galactic cartesian velocities $\bm{V}_\mathrm{car}=(V_\mathrm{x},V_\mathrm{y},V_\mathrm{z})$ to the the observed velocities, while $\mathbf{R}_\mathrm{s,sph}$ and $\mathbf{R}_\mathrm{s,cyl}$ are the rotation matrix to pass from Galactic cartesian velocities to Galactic spherical and cylidrincal velocities, respectively. 
The matrix $\mathbf{R}_\mathrm{c}$  is defined as 
\begin{equation}
\mathbf{R}_\mathrm{c} =
\begin{bmatrix}
\cos b \cos \ell & \cos b \sin \ell & \sin b  \\
-\sin \ell  & \cos \ell & 0 \\
- \sin b \cos \ell & -\sin b \sin \ell & \cos b
\end{bmatrix},
\end{equation}
while the matrices $\mathbf{R}_\mathrm{s}$ are defined as
\begin{equation}
\mathbf{R}_\mathrm{s,sph} =
\begin{bmatrix}
\Gamma \cos \theta \cos \phi & -\Gamma\sin \theta \cos \phi & -\Gamma \sin \phi  \\
\cos \theta \sin \phi  & -\sin \theta \sin \phi & - \cos \theta \\
\sin \theta & \cos \theta & 0
\end{bmatrix}
\end{equation} 
and
\begin{equation}
\mathbf{R}_\mathrm{s,cyl} =
\begin{bmatrix}
\Gamma  \cos \phi & 0 & -\Gamma \sin \phi  \\
\sin \phi  & 0 & - \cos \phi \\
0 & 1 & 0
\end{bmatrix}.
\end{equation} 
The factor $\Gamma$ is equal to 1 for a right-handed Galactocentric frame of reference or to -1 for a left-handed Galactocentric frame of reference (as the one  used in this work). The angular coordinates $\theta$ and $\phi$ are the zenithal and azimuthal angle respectively, while $b$ and $\ell$ are the Galactic sky coordinates (see Sec.\ \ref{sec:dist_estimate}).

\bsp	
\label{lastpage}
\end{document}